\documentclass[aps,amsmath,twocolumn,amssymb,floatfixng,showpacs,superscriptaddress,footinbib]{revtex4-1}
\pdfoutput=1
\usepackage[dvips]{graphics}
\usepackage{bm}
\usepackage{float}
\usepackage{epsfig}
\usepackage{enumerate}
\usepackage{subfigure}
\usepackage{amsmath}
\usepackage{color}
\usepackage{braket}
\usepackage{graphicx}
\usepackage{float}
\usepackage[colorlinks=true,linktoc=page,linkcolor=red,citecolor=blue,urlcolor=magenta]{hyperref}

\newcommand\bea{\begin{eqnarray}}
\newcommand\eea{\end{eqnarray}}
\newcommand\beq{\begin{equation}}  
\newcommand\eeq{\end{equation}}

\begin{document} 

\title{Adiabatic charge transport through non-Bloch bands} 
 \author{Dharana Joshi}
\affiliation{Department of Physics, BITS Pilani-Hyderabad Campus, Telangana 500078, India}
\author{Tanay Nag}
\email{tanay.nag@hyderabad.bits-pilani.ac.in}
\affiliation{Department of Physics, BITS Pilani-Hyderabad Campus, Telangana 500078, India}


\begin{abstract}
We explore the non-reciprocal intracell hopping mediated non-Hermitian topological phases of an extended Su-Schrieffer-Heeger model hosting second-nearest-neighbour hopping. We microscopically analyze the phase boundaries using the non-Bloch momentum while the off-critical (critical) phases are directly associated with the gapped (gapless) nature of the non-Bloch bands that we derive from the characteristic equation using the gauge freedom. The non-Bloch momentum accurately reflects the bulk boundary correspondence (BBC) explaining the winding number profile under open boundary conditions. We examine the 
adiabatic dynamics to promote the concept of adiabatic charge transport in a non-Hermitian scenario justifying the BBC in spatio-temporal Bott index and non-Bloch Chern number. Once the  non-Bloch bands experience no (a) gap-closing during the evolution of time, quantized flow of is preserved (broken). Our study systematically unifies the concept of non-Bloch bands for both static and driven situations.

\end{abstract}

\maketitle


\textcolor{blue}{\textit{Introduction:}}
In recent times, 
the non-Hermitian (NH) system  
acquires a lot of attention due to their ability of mimicking   an open quantum system effectively \cite{rotter2009non,Niu23,roccati2022non}. The environmental effects such as  gain and loss, are already embedded in   metamaterials \cite{ghatak2020observation} where the mean energy is not strictly conserved extending the periphery of quantum materials beyond Hermiticity \cite{wang2023non,lu2025non,Zhou20}. 
The topological phases, being an important upshot of quantum materials, are extensively studied in the context of Hermitian models, and they have been recently extended to NH models \cite{Okuma_2023,Gong18,Liang2013,Leykam_2017,Martinez2018,Xu_2016,Doppler_2016,Ghatak_2019,Banerjee_2023,Mondal22,Mondal23,mondal2024persistent}. The degeneracies in Hermitian systems transform into exceptional points for NH system where the complex energy spectrum is more prone to gap closing \cite{Bergholtz21,Budich20,Bergholtz19}. Surprisingly, the finite energy modes exhibit strong boundary localization, otherwise located in the bulk of a Hermitian topological model, which is known as NH skin effect being a hallmark signature of NH topological system \cite{Okuma_2023,Gong2017,Lee_2016,gohsrich2024,Kawabata2019,Wang_2024,Denner_2021,Bao2024,Han_2021,Kawabata20,zhang2021observation,Ma24,Okugawa20,lin2023topological,Ghosh22,Ghosh2024}. The zero-energy boundary modes, obtained from open boundary condition (OBC), are characterized by bulk invariant, computed from periodic boundary condition (PBC), in Hermitian systems, known as     bulk-boundary correspondence (BBC) which is apparently broken in NH topological model owing to the presence of non-Bloch momentum \cite{Xiong_2018,Koch_2020,Song2019,Yao2018,Kawabata_2020,Yokomizo2019,He_2021,Lin2021,Ghosh2024,zhang2022,Tang2020,Song2019,He_2021,Kivelson82,hamano2024}.

The Su-Schrieffer-Heeger (SSH)  model and its variants are found to be a general testbed to investigate various Hermitian and NH topological effects due to its simple structure \cite{Anastasiadis22,Verma24,du2024one,dharana2025,rajbongshi2025topological,asboth2016short,Maffei_2018,Perez2019,Du_2024,Feng2022,Halder_2022,Lieu_2018,Nehra_2022,Lieu18,Yin18,PhysRevB.100.045141}. A cyclic and slow  variation of system parameters over time, allowing adiabaticity to hold, can result in the quantized transport of charge across the dimension of the system.  This is directly related to the bulk topological invariants enabling the  charge pumping a powerful probe of topological phases \cite{Citro_2023,NIU91,shun2018topological,Padhan2024} which has NH analog as well \cite{Kumar_2025,Zhang_2024}. 
While much has been studied on the different types of NH terms in SSH and SSH-like models and their effects \cite{Han_2021,halder2022properties,he2020non,Nehra_2022,verma2024non,verma2024topological}, the microscopic structure of the non-Bloch form of momentum remains mostly unexplored \cite{Yokomizo2019,Yao18,Yang20,He_2021}. Considering an extended SSH model with second-nearest-neighbour hopping, we would like to understand the NH topological phase diagram using the non-Bloch momentum and their effective band structures in  static as well as adiabatically driven cases. To be precise, 
the main questions that we address here are the following: How to understand an extended critical region  from non-Bloch momentum? How does non-Bloch momentum establish BBC in the context of adiabatic charge transport?

Our study on the extended SSH model with non-reciprocal hopping reveals a complex structure of the non-Bloch momentum that varies significantly with parameters without having any Hermitian analogue. The characteristics equation, consisting of energy and complex momentum, in conjunction with gauge freedom, is solved self-consistently to obtain the continuous energy bands associated with a closed loop of non-Bloch momentum. The 
gapped and gapless phases, indicated by the  OBC analysis, are  accurately captured by the non-Bloch band dispersion in PBC irrespective of their topological nature while topology is apprehended through BBC between winding number computed using bi-orthogonalized open-bulk and non-Bloch bulk bands. We study the adiabatic dynamics of the model where the time-evolution of non-Bloch bands can capture the gap-closing in the driven scenario as well. The  BBC, carrying the  signature of quantized transport,  is satisfied with the spatio-temporal Bott index  and the non-Bloch Chern number. Our study hence justifies the emergence non-Bloch momentum for a driven NH system as well.


\begin{figure}
    \centering
    \includegraphics[width=1\linewidth]{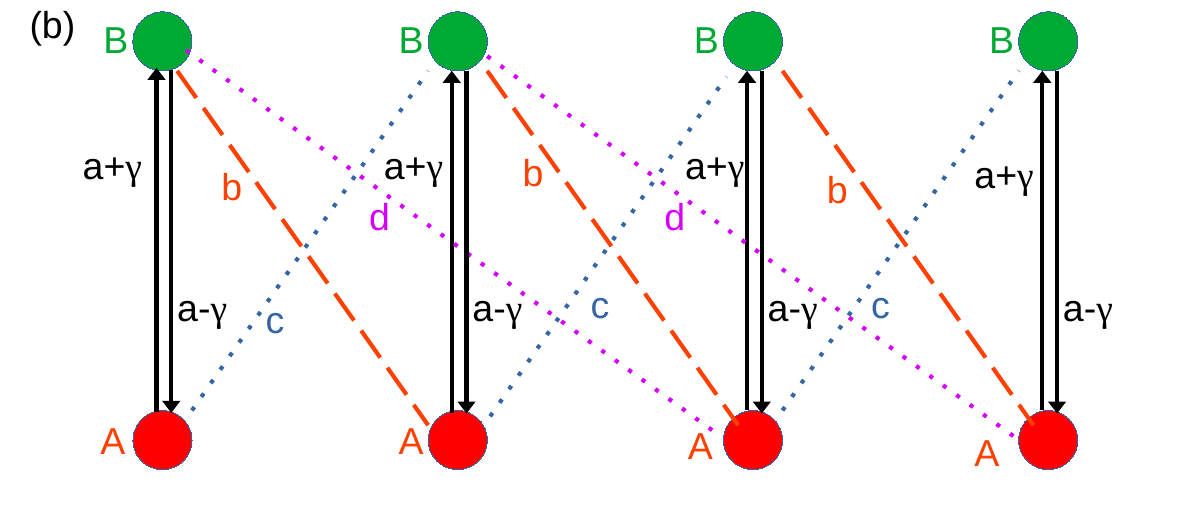}
    \caption{Schematic diagram of the SSHLR model with two sub-lattices $A$ and $B$ per unit cell. The inter(intra)-cell hoppings $b,c$, and $d$ are designated by dashed and dotted (solid) lines where $\gamma$ introduces non-reciprocity in the  intra-cell hopping.}
    \label{fig:Figmodel}
\end{figure}

\begin{figure*}
    \centering
    \includegraphics[width=1\linewidth]{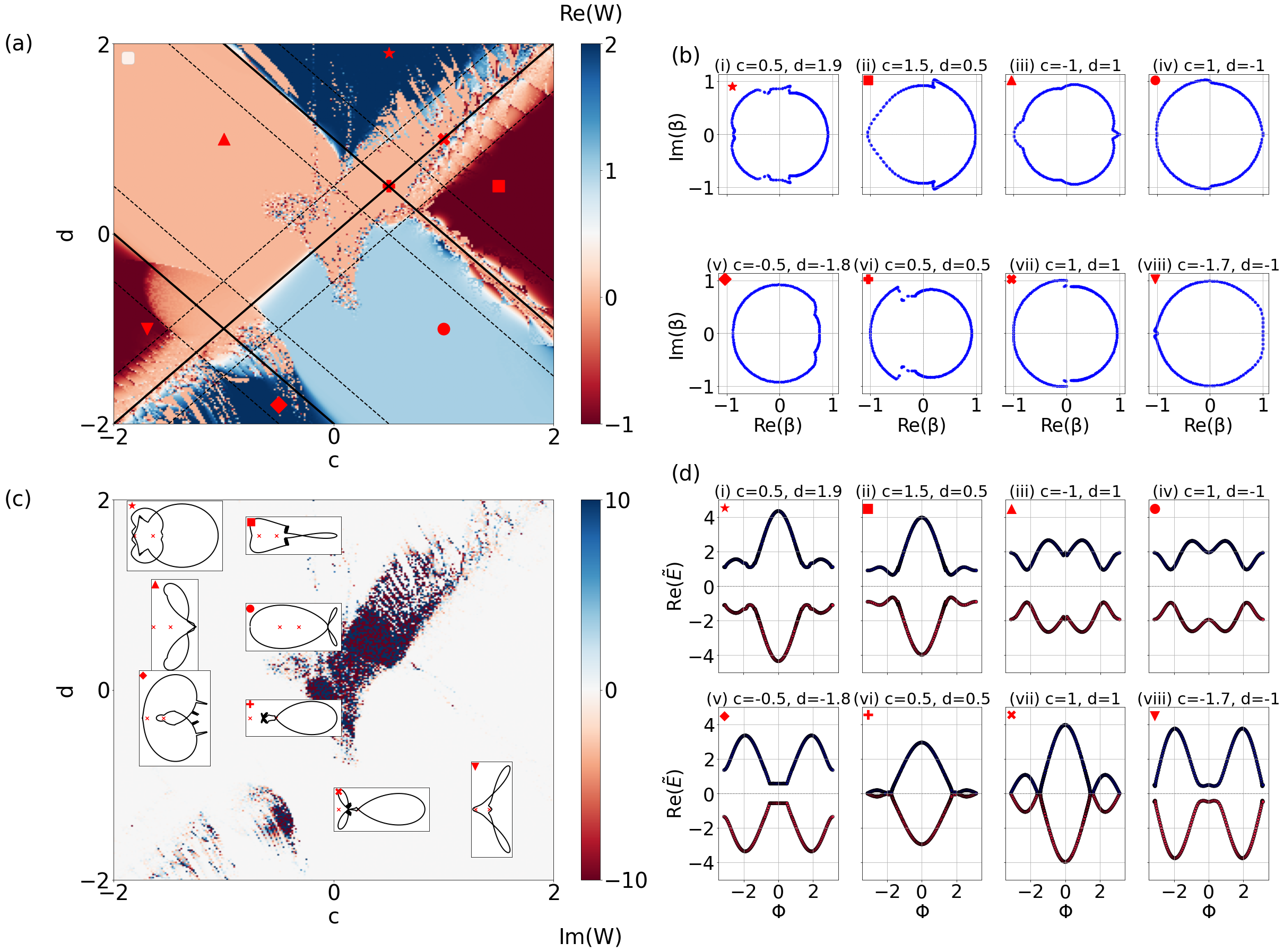}
    \caption{We show the phase diagram of open-bulk winding number ${\rm Re}[W]$ (color bar), using  Eq. (\ref{eq:rwind}), in (a) as a function of $c$ and $d$. The solid (dashed) lines represent the Hermitian (NH) phase boundaries for $k=0$, $\pi$ and $2\pi/3$. We consider  $a=b=1$, $\gamma=0.5$, $L=70, l=15$.  (b) shows the GBZs in complex $\beta$ plane, corresponding to each parameter point as marked by red symbols $\star$, $\square$, $\triangle$, $\circ$, $\diamond$,  $\triangledown$, $+$, and $\times$ in (a). We depict  ${\rm Im}[W]$ in (c) where the extended critical regions acquire  finite ${\rm Im}[W]$. The insets in (c) show the parametric plot in $d^x_R(\beta)$-$d^y_R(\beta)$ plane for different regions of the phase diagram as marked by red symbols in (a).   The  continuous non-Bloch energy spectra ${\tilde E}$ with polar angle $\phi=\tan^{-1}({\rm Im[\beta]/{\rm Re[\beta]}})$ are shown in  (d) for the above marked points, see SM \cite{supp} for more details. }
    \label{fig:fig2}
\end{figure*}
 

\textcolor{blue}{\textit{Model:}}
We consider NH  SSH long-range (SSHLR) model by employing the asymmetrized hopping parameter $\gamma$ into the intra-cell hopping parameter of the underlying Hermitian model \cite{Maffei_2018,dharana2025}. To be precise, $A_n \rightarrow B_n$ ($B_n \rightarrow A_n$) hopping is modified from $a$ to $a+\gamma$ ($a-\gamma$), ensuring the non-reciprocity in the hopping amplitudes, as shown in the Fig \ref{fig:Figmodel}. The momentum space Hamiltonian, describing the NH SSHLR model in the basis $(C^{\dagger}_{A,k}, C^{\dagger}_{B,k})$, is given by
 \begin{equation}
H(k)=\bm{d}\cdot {\bm \sigma}=
H_{\mathcal H}(k) + i \gamma \sigma_y
\label{eq:sshlr1}  
\end{equation}
where ${\bm d}={\bm d}_R+ i {\bm d}_I= (d^x_R+i d^x_I, d^y_R+i d^y_I, d^z_R+i d^z_I) $, ${\bm \sigma}=(\sigma_x,\sigma_y,\sigma_z)$, 
the Hermitian part is given by $H_{\mathcal H}(k)= \bm{d}_R(k)\cdot {\bm \sigma}=(a+b\cos k+ c \cos k + d\cos 2k) \sigma_x + (b\sin k - c \sin k +d \sin 2k)\sigma_y$ and NH part is written as ${\bm d}_I=(0,\gamma,0)$ with  $d^y_I=\gamma$. This model respects chiral symmetry $\Gamma=\sigma_z$, such that $\Gamma H(k) \Gamma^{-1}=-H(k)$. For the NH case, the  degeneracies are abundant when $ d^2_R=d^2_I$ and ${\bm d}_R\cdot {\bm d}_I=0$
are simultaneously satisfied. This leads to the following phase boundaries $c=-2-d \pm \gamma$, $c=d \pm \gamma$, and $c=1-d \pm \gamma$ for $k=0$, $\pi$ and $2\pi/3$, respectively (keeping $a = b = 1$), see Fig. \ref{fig:fig2} (a). Note that the Hermitian phase boundaries are correctly reproduced when $\gamma=0$ \cite{dharana2025}.

As anticipated from the  above momentum space analysis, NH phase boundaries get shifted by an amount $\pm \gamma$ with respect to the Hermitian phase boundaries. 
These new phase boundaries  often referred to as exceptional lines \cite{Kumar_2025,Kawabata_2020}, however, one needs to check whether such critical lines are supported by the Hamiltonian under OBC. 
In order to verify the BBC, we compute the real space open-bulk  bi-orthogonalized 
winding number $W$ under OBC as given below \cite{Song2019,He_2021,dharana2025}
\begin{equation}
    W=\frac{1}{2L}\mbox{Tr}'\Big(\tilde{\Gamma}Q[Q,\tilde{X}]\Big).
\label{eq:rwind}
\end{equation}
Here, $\tilde{X}$ and $\tilde{\Gamma}$ are the position and chiral symmetry matrices, respectively for the lattice.  $Q$ is the flattened Hamiltonian.  ${\rm Tr}'$ represents partial trace excluding the boundary sites, see supplementary material (SM) \cite{supp} for more details.
Interestingly, we find that $\pm \gamma$ shift of the Hermitian boundaries are not in accordance with the topological phase diagram, obtained from the real part of $W$ in $c$-$d$ plane, as shown in Fig. \ref{fig:fig2}(a).   The  
stable (fluctuating) behavior of winding number signifies the existence of gapped (gappless) phases. 
To further emphasize this aspect, we plot the imaginary part of $W$ in Fig. \ref{fig:fig2}(c) which is directly correlated with the fluctuations in Re[$W$] around the Hermitian phase boundaries. This instigated us to investigate the non-Bloch momentum $\beta$ that  we explore below.

\textcolor{blue}{\textit{Non-Bloch momentum and bulk invariant:}}
One can introduce a generalized Bloch Hamiltonian $H(\beta)$ for NH model by substituting $e^{ik} \rightarrow \beta$ in the momentum-space Hamiltonian Eq. (\ref{eq:sshlr1}); $\beta$ is a complex quantity.  One needs to solve the characteristic equation det$[H(\beta)-E]=0$, leading to a sixth degree polynomial equation $f(\beta)+E^2=0$ in $\beta$, to obtain $\beta$ for a given energy $E$. 
Importantly,  exploiting the  gauge freedom  $f(\beta)=f(\beta e^{i\theta})$ along with polynomial equation, one can numerically obtain $\beta $ and $E= \tilde{E}$ for a given value of $\theta$  with $\theta \in (0,2\pi) $\cite{Yokomizo2019} while analytical roots of $\beta$  are independent of $\theta$ and $E$. The primary roots $\beta_p$ constitute the  auxiliary generalized Brillouin zone (aGBZ) which is nested due to the multi-valued behavior of the roots for a given $\theta$. A self-consistent procedure, using the above two equations, is employed to obtain a single-valued $|\beta|$ and $\tilde{E}$  which are uniquely associated with $\theta$ resulting in the formation of generalized BZ (GBZ), see SM for more details \cite{supp}. In this procedure,  one obtains six secondary roots $\beta_s$, satisfying $| \beta_{s,1} | \le |\beta_{s,2}|.........|\beta_{s,5}| \le|\beta_{s,6}|$. The convergence is obtained when there exist middle two roots sharing the same absolute value in the above set and simultaneously they yield continuous real energy bands with ${\rm Re}[\tilde{E}]$  as a function of $\theta$ \cite{Yokomizo2019}. Note that all values of $\theta$ do not necessarily contribute in forming the GBZ.

We demonstrate different GBZ in the complex plane of $\beta$  where we make eight distinct choices of parameters $(c,d)$ over the phase diagram to investigate the effect of topological phase on the shape of the GBZ, see Fig. \ref{fig:fig2}(b). We find kink-like pattern emerges around ${\rm Re}[\beta]=0$ as well as around ${\rm Im}[\beta]=0$ indicating towards the fact that long-range hopping causes deviation in GBZ from a  circular one. It is noteworthy that there exist only two  $\beta$ roots given by $|\beta_{s1}|=|\beta_{s2}|= \sqrt{(a-\gamma)/(a+\gamma)}<1$ irrespective of the choice of $\theta$, leading to a circular GBZ in NH SSH model without long range hopping i.e., $c=d=0$ \cite{He_2021,Yao2018}. These kink-like structure in GBZ is more evident for the critical phases, demarcated by $\times$ and $+$ marks around $c=d$ line, where the ${\rm Re}[W]$ fluctuates and  ${\rm Im}[W]$ shows non-zero values, see Figs. \ref{fig:fig2}(a,c).

Taking the non-Bloch momentum analysis further,  the polar angle $\phi=\tan^{-1}({\rm Im[\beta]/{\rm Re[\beta]}}) \in [-\pi,\pi]$ can be obtained from GBZ as shown in Fig. \ref{fig:fig2}(b) while $\theta$ acts as the probe to obtain the correct set of $\beta$ and $|\beta|$ denotes the radial distance at a given $\phi$. Therefore, the Hermitian BZ  on unit circle, designated by $e^{ik}$, gets converted into a GBZ on closed complex loop, characterized by $e^{ik} \to \beta=|\beta| e^{i\phi}$ where $|\beta|$ generically depends on $\phi$.
We illustrate the winding of non-Bloch vectors from Eq. (\ref{eq:sshlr1}), $d^x_R(\beta)$ and $d^y_R(\beta)$ around each other  while traversing across GBZ with $\phi \in [-\pi,\pi]$ in the inset of Fig. \ref{fig:fig2}(c) for 
different values of $(c,d)$ as symbolized in the Fig. \ref{fig:fig2}(a). The topology is determined if at least one out of the two exceptional points $\big(d^x_R(\beta),d^y_R(\beta)\big)=(\pm \gamma,0)$ is enclosed and the number of winding dictates the winding number.  The critical phase is captured when the parametric loop touches any one of the two exceptional point as seen for $+$ and $\times$ symbols. This closely resembles the winding of the Bloch vectors around the origin  while traversing across BZ of a Hermitian model to determine its topology.

Importantly, the effective energy ${\rm Re}[{\tilde E}]$, associated with the converged $\beta$ forming GBZ, of the non-Bloch bands can be linked with $\phi$ that  plays the role of an effective momentum, see Fig. \ref{fig:fig2}(d).  The off-critical (critical)  phases exhibit  a gap (gap-closing) between valence and conduction bands as shown with  $\star$, $\square$, $\triangle$, $\circ$, $\diamond$, and $\triangledown$ ($+$, and $\times$)  symbols. This non-Bloch band structure respects the chiral symmetry as the energy of the valence band is exactly opposite to that of the conduction band for a given $\phi$.
The gap observed is only topological in nature when ${\rm Re}[W]\ne 0$ ${\rm Im}[W]=0$, while gapless region is associated with  ${\rm Re}[W]=0$ ${\rm Im}[W] \ne 0$. Hence the critical behavior correctly justifies the finite ${\rm Im}[W]$ behavior around the Hermitian phase boundaries.

 \begin{figure}
    \centering
    \includegraphics[width=1\linewidth]{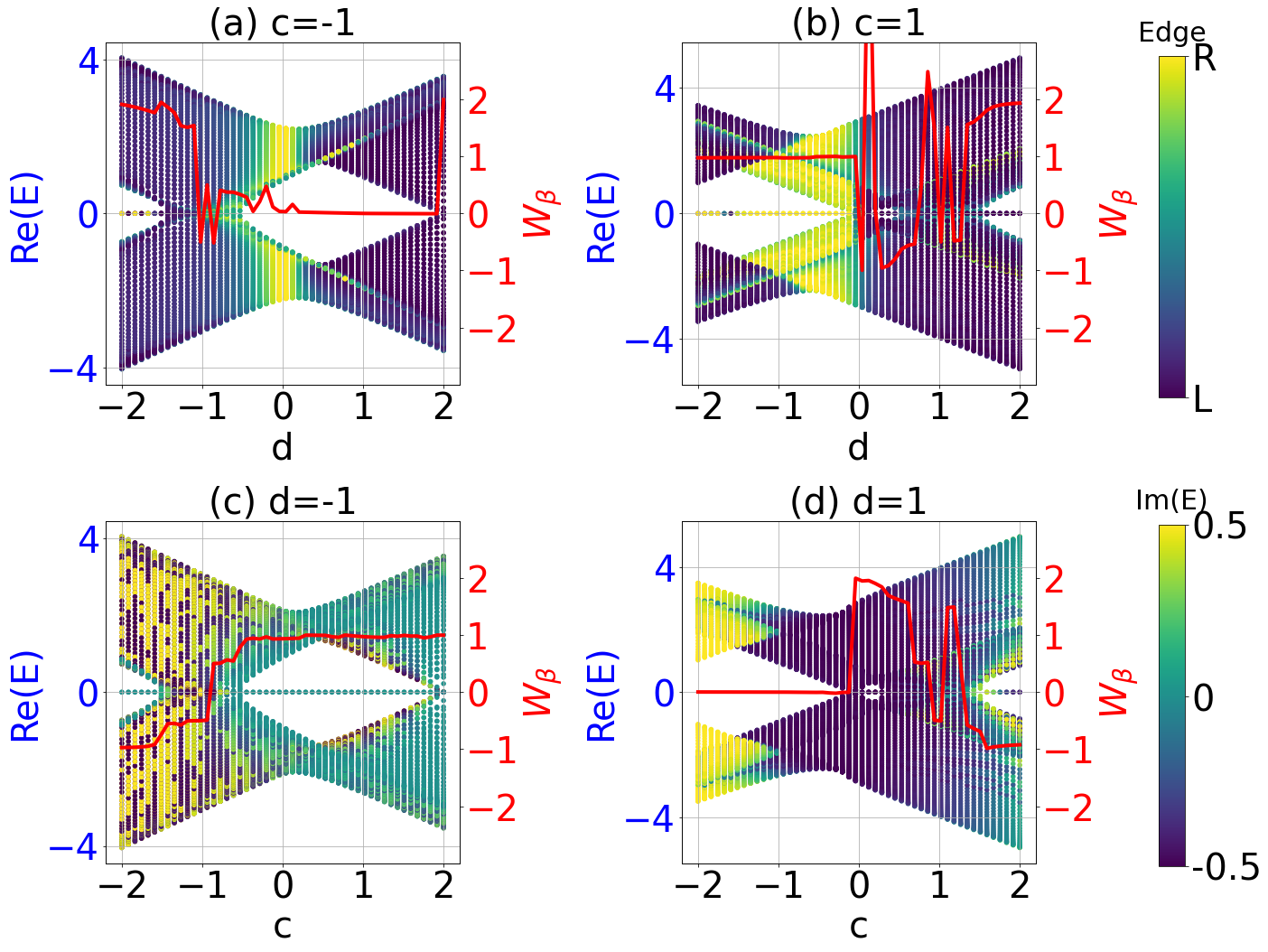}
    \caption{We show the evolution of energy levels (left axis ${\rm Re}[E]$), computed from real space NH SSHLR model under OBC, and winding number, obtained from bulk non-Bloch model $H(\beta)$ (right axis $W_{\beta}$), with $d$ and $c$ in (a,b) and (c,d), respectively. We consider $c=-1$, $c=1$, $d=-1$ and $d=1$ in (a,b,c,d), respectively. For $W_{\beta}$, we integrate over 200 values of $\phi \in[0, 2 \pi]$ values and system size is $N=100$. We color-code (a,b,d) with the average position i.e., blue and yellow for left (L) and right (R) edges, respectively, of the individual energy level and (c) with the ${\rm Im}[E]$.   }
    \label{fig:fig3}
\end{figure}


We now investigate the BBC using the non-Bloch momentum $\beta$ that constitutes GBZ. We compute the   bulk bi-orthogonalized winding number $W_{\beta}$
given by 
\begin{equation}
    W_{\beta}=\frac{1}{2\pi i} \int_0^{2\pi} d\phi \, \partial_\phi \log\big[\det(q)\big] 
\end{equation}
where $q$ is obtained from the anti-diagonal part of the flattened Hamiltonian $Q$ which is obtained from  $H(\beta)$ (Eq. (\ref{eq:sshlr1})). Here $\phi$ is a periodic variable. We show the evolution of real-space OBC energy spectrum alongside with $W_{\beta}$ as a function of a parameter to establish the  non-Bloch BBC that is otherwise not recovered with the Bloch momentum $k$, see Figs. \ref{fig:fig3} (a,b) and (c,d) for variation with $d$ and $c$, respectively.   These results clearly reveal the direct correspondence between the existence (number) of zero-energy modes and quantized (non-zero) values of $W_{\beta}$. The phase boundaries are correctly mimicked, irrespective of the phases on either side. Most importantly, we find critical phases where a highly degenerate gapless region is directly associated with a non-quantized and fluctuating nature in $W_{\beta}$. The color-coded energy spectrum in terms of the average position straightaway brings the existence of NH skin effect in our model, see Figs. \ref{fig:fig3} (a,b,d). The imaginary part of the energy survives in the  critical region exhibiting a complex structure at ${\rm Re}[E]=0$ line whereas for topological phases, the earlier always vanishes, see Figs. \ref{fig:fig3} (c). Therefore, after correctly identifying the non-Bloch momentum $|\beta| e^{i\phi}$, the topological phases observed under OBC can be appropriately understood from the bulk invariant.

\begin{figure*}
    \centering
    \includegraphics[width=1\linewidth]{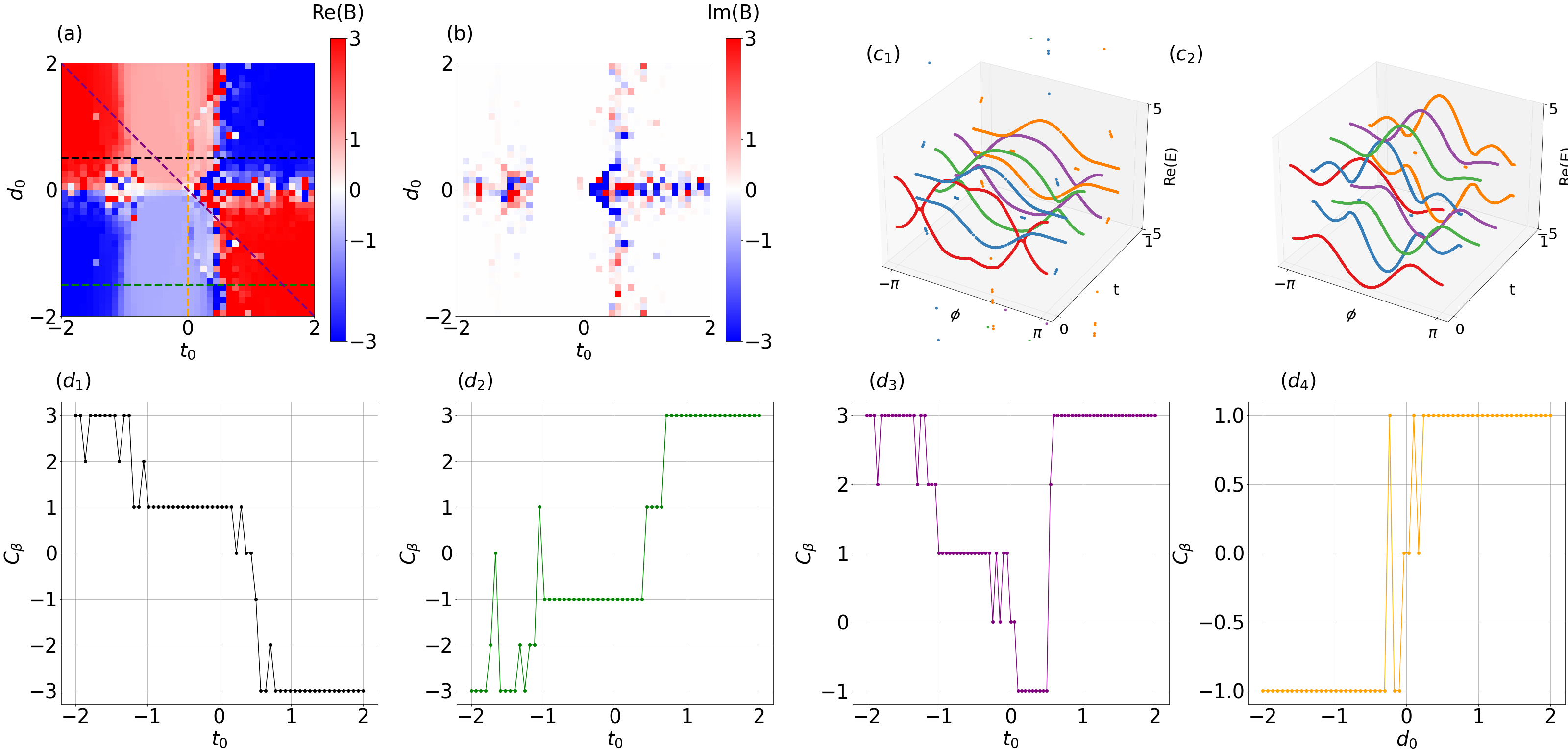}
    \caption{We show the topological phase diagram, obtained using open-bulk Bott index ${\rm Re}[B]$ given by Eq.(\ref{eq:bott_obc}) (color bar), over $d_0$-$t_0$ plane in (a). We consider  $a = b = h_0 = T = 1$, and  $L_x = L_y = 50$. We depict  ${\rm Im}[B]$ to highlight the extended critical phase in (b). 
    We show the time evolution of the effective dispersion $\tilde {E}$-$\phi$ associated with non-Bloch bands  in (c1,c2) for $(t_0,d_0)=(0.5,1)$ and $(1.5,1.5)$, respectively. We demonstrate the variation of non-Bloch  Chern number $C_\beta$, associated with the ground state, in $(d_1,d_2,d_3,d_4)$,   
    corresponding to the lines $d_0=-1.5$ (black), $d_0=0.5$ (green), $d_0=-t_0$ (purple), $t_0=0$ (yellow) in (a), respectively.    }
    \label{fig:fig4}
\end{figure*}


\textcolor{blue}{\textit{Adibatic transport and time-dependent non-Bloch momentum:}}
To this end, we consider time-dependent long range terms in $H_{\mathcal H}$ (Eq. (\ref{eq:sshlr1})) along  with a chiral symmetry breaking term  that $H(k,t)=h(t) \sigma_z + (a+b\cos k+ c(t) \cos k + d(t)\cos 2k) \sigma_x + (b\sin k - c(t) \sin k + d(t) \sin 2k)\sigma_y + i \gamma \sigma_y$; $c(t)=t_0-d_0 \cos (2\pi t/T)$, $d(t)=t_0+d_{0} \cos (2\pi t/T)$, and $h(t)=h_0 \sin (2\pi t/T)$ \cite{Citro_2023,NIU91,dharana2025}. Note that we consider the dynamics under adiabatic limit with $T=2\pi/\omega \gg \Delta^{-1}$ where $\Delta$ ($\omega$) is the energy gap (drive frequency), defining the internal  relaxation time (external time period $T$) of the model. One can easily check that $\Delta \ne 0,~\forall ~t \in [0,T]$ as either $d_z(t)$ is non-zero or $d_{x,y}(t)$.  Therefore, the   system always follows the
instantaneous ground state while the time is treated as a parameter. In other words, the quasi-states and  quasi-energies, obtained using a full Floquet theory, is directly computed  from the eigenstates of
$H(t=T)$ and $(1/T)\int_0^T E(t) dt$, respectively, where $E(t)$ is the eigenvalue of $H(t)$ \cite{Nag15}.  This adiabatic limit allows us to probe the dynamical phase diagram using the 2D invariant spatio-temporal Bott index and  Chern number in real and momentum spaces, respectively, with the conversion $(k,t)\equiv (k_x,k_y) \to (x,y)$ \cite{dharana2025}. We below study the fate of quantized charge pumping in the NH scenario.

We now compute the real space open-bulk bi-orthogonalized  Bott index $B$  under OBC   \cite{He_2021, Song2019,bellissard1995,Prodan_2010,Bianco11,Loring_2010,hastings2010,Yoshii_2021,loring2019,Toniolo2017,Toniolo2018},  defined as 
\begin{equation}
    B=\frac{2\pi}{\tilde{L}_x \tilde{L}_y} {\rm Im}\big({\rm Tr'}[PxP,PyP]\big)
    \label{eq:bott_obc}
\end{equation}
where $P$ is the projector onto the occupied states, see SM \cite{supp} for more details. Note that  without ${\rm Tr}'$, the imaginary part signifies the local Chern marker in $(x,y)$-grid. We demonstrate the Bott index namely, ${\rm Re}[B]$ in $d_0$-$t_0$ plane, see Fig \ref{fig:fig4} (a), where we find NH phase boundaries follow Hermitian phase boundaries $d_0=0$, $t_0=-1$ and $1/2$ except 
strong fluctuations around the intersection of two boundaries around $(t_0,d_0)=(-1,0)$ and $(1/2,0)$.  However, the fluctuations persist  around $t_0=1/2$, and $d_0=0$  phase boundaries. The quantized nature of the charge pumping is completely disrupted in these critical regions. 
Likewise the static case, NH introduces critical regions, associated with ${\rm Im}[B]$ as shown in Fig. \ref{fig:fig4}(b), around the Hermitian phase boundaries leading to the fact that
non-reciprocity in the hopping causes a shrink in the underlying topological Hermitian phases. Away from the boundaries, deep inside the topological phases, Bott index accurately characterizes the quantized charge pumping except for a few defective points where finite size effects dominate.

We would now like to connect the above OBC behavior with the PBC bulk invariant to establish the dynamic  BBC.   Consequently, we explore the time evolution of GBZ, see SM \cite{supp} for more details.  We compute GBZ following the method discussed above, at each instant of time $t \in [0,T]$, leading to the time evolution of the non-Bloch bands. Interestingly, non-Bloch bands show degeneracy in more than one time slice justifying the irregularity in ${\rm Re}[B]$ for the 
critical regions where charge pumping no longer remains quantized, see Fig. \ref{fig:fig4} $(c_1)$.  For topological regions, one can find that non-Bloch bands are always gapped out during the course of adiabatic evolution and thus ensuring the quantized charge pumping, see Fig. \ref{fig:fig4} $(c_2)$.

The topological nature of the gap is examined using the non-Bloch Chern number for the \(n\)th band that can be expressed as \cite{fukui2005chern}
$C^n_{\beta} = \frac{1}{2\pi i} \int_{0}^{2\pi} \int_{0}^{T} d\phi \, dt \, F_{12,n}(\phi,t)$
where $F_{12,n}(\phi,t) = \partial_1 A_{2,n}(\phi,t) - \partial_2 A_{1,n}(\phi,t)$ ($1,2=\phi,t$) denotes the Berry curvature and $A_{\mu,n}(\phi,t) = \bra{n_L(\phi,t)} \partial_\mu \ket{n_R(\phi,t)}$ represents the bi-orthogonalized Abelian Berry connection. Here, $\ket{n_{L,R} (\phi,t)}$  are left and right eigenvector of the time-dependent Hamiltonian $H(\beta,t)$. Notice that we have used an interpolation scheme to make $(\phi,t)$ uniform across the parameter space, see SM \cite{supp} for more details. We choose four paths, colored by dashed black, green, violet and yellow lines, on the $d_0$-$t_0$ plane, to illustrate the variation of non-Bloch Chern number in Figs. \ref{fig:fig4} $(d_1), (d_2), (d_3)$ and  $(d_4)$, respectively. Remarkably, we find that the non-Bloch bulk Chern number perfectly mimics the behavior of the open-bulk spatio-temporal Bott index except for a few isolated points. The quantized nature of the Chern number verifies the existence of pumped charge in the NH context.

\textcolor{blue}{\textit{Discussion:}}
In this work, non-reciprocal hopping causes the bulk states to move one end of the lattice, enforcing the NH skin effects that we see in Fig. \ref{fig:fig3}. The on-site gain and loss are also found to be responsible to generate the NH skin effect \cite{Lai25,vs7x-clqd}. The NH skin effect is microscopically caused by the non-Bloch momentum where the real part of $\beta$ yields an exponential decay of the finite-energy states in the bulk of the lattice in addition to the zero-energy edge modes. The critical regions with a macroscopic number of degeneracies are mainly caused by the second-nearest-neighbour hopping while the first-nearest-neighbour hopping allows for a lesser number of degeneracies within the extended critical regions \cite{Ghosh24b,han2021topological}. The number of critical momentum increases with the range of hopping in Hermitian model indicating towards the fact that non-Bloch bands touch each other at 
different values of $\phi$ other than $0$ or $\pi$ in NH model as noticed in Fig. \ref{fig:fig2} (d). 
It is noteworthy that non-reciprocal hopping steers to shrink topological phases for static and adiabatic drive both  for the parameter $a=b=1$ which represents a critical point in the  Hermitian SSH model with first-nearest-neighbour hopping. It is therefore, possible to obtain NH-mediated topological phase for $a \ne b$ where the same non-Bloch theory would work to establish the BBC.

Note that the NH topological phases  are protected by  ramified  time-reversal symmetry  and particle-hole symmetry, generated by unitary operators  ${\mathcal T}=I$ and ${\mathcal P}=\sigma_z$, respectively   such that ${\mathcal T} H^*(k) {\mathcal T}^{-1}=H(-k)$ and   ${\mathcal P} H^*(k) {\mathcal P}^{-1}=-H(-k)$ \cite{Kawabata19}.
The non-Bloch spectrum in Fig. \ref{fig:fig2} (d) respects the above 
symmetries in terms of $\phi$ which plays the role of $k$ such that $\tilde{E}(\phi)=-\tilde{E}(-\phi)$, $\tilde{E}(\phi)=\tilde{E}(-\phi)$. 
In the case of adiabatic drive, the time-evolved non-Bloch bands break the $\Gamma$, ${\mathcal P}$ and ${\mathcal T}$ symmetries resulting in the emergent NH 2D Chern phases. In this case,  the non-Bloch winding over $d^x_R(\beta, t)$-$d^y_R (\beta, t)$ plane occurs with time justifying charge transport. To be precise, for $t=0, T/2$, $d^z_R(t)=0$ causing the static non-Bloch winding analysis applicable, see SM \cite{supp} for more details. Given the experimental advancements in NH physics \cite{li2024observation,lin2022experimental,li2019observation,PhysRevLett.123.165701} and  SSH-type topological lattices \cite{meier2016observation,lee2018topolectrical,atala2013direct}, we anticipate that engineered optical-lattice architectures and topolectrical circuit networks \cite{goldman2016topological,PhysRevResearch.3.023056,yuan2023non,wang2022non} constitute promising platforms for implementing and probing the predicted phenomena in our work.
Note that adiabatic particle transport has been experimentally studied through the  center-of-mass displacement and light intensity distribution in  cold atom systems and photonic waveguide arrays, respectively \cite{Lohse_2015,Nakajima_2016,lohse2018exploring, cerjan2020}. The experimental realizations in non-Hermitian systems using optical lattices and photonic waveguides with controlled dissipation \cite{PhysRevLett.124.250402,zhao2025two,PhysRevLett.121.150403,xiao2025non} enable the manifestation of the biorthogonal Berry connection \cite{Qin2024,yin2025} through the pumped charge, which can be detected via the observables discussed above. In solid-state systems, our proposal can be validated when the potential difference between left and right leads is slowly varied within the bulk gap of the central system to transfer charge across it in a quantized manner \cite{PhysRevB.106.045417,PhysRevB.110.045138,doi:10.1126/science.aar3766}.

\textcolor{blue}{\textit{Conclusion:}}
We consider an extended SSH model with second-nearest-neighbour intercell hopping namely, SSHLR model in the presence of a non-reciprocal intracell  hopping to study the emergence of NH topological phases in statics as well as under adiabatic drive. To begin with, we show that the non-Bloch momentum acquires complicated closed loops depending upon the choice of parameter which does not have any Hermitian analog. By solving the characteristic equation, and employing gauge freedom,  we obtain the  non-trivial winding of the non-Bloch bands and their effective energies. The critical and  off-critical phases are properly connected to  gapless and  gapped nature of the non-Bloch spectrum, respectively.  The topology of the gap is directly obtained from the quantized open-bulk invariant as well as non-Bloch bulk invariant ensuring the BBC  nicely.  On the other hand,  the non-quantized  irregular behavior of the topological indices is directly connected with the gapless nature of the non-Bloch spectrum.
We next invoke time to analyze the adiabatic charge transport in a NH SSHLR   model. We similarly obtain the quantized charge transport is directly related to the gapped structure of the non-Bloch bands with time while non-quantized charge transport is caused by the 
gap-closing in  non-Bloch bands with time. We study bi-orthogonalized open-bulk spatio-temporal Bott index to characterize these topological phases. The BBC is perfectly respected when we combine the above with the non-Bloch Chern number. Therefore, our study reveals the non-Bloch analogy of NH topology for static and driven system that could potentially lead to similar future studies on Floquet topological insulator and topological superconductor models.

\textcolor{blue}{\textit{Acknowledgement:}}
We sincerely thank Sanchayan Banerjee and Tapan Mishra for  useful discussions on the numerical roots of non-Bloch momentum.   T.N. acknowledges the
NFSG ``NFSG/HYD/2023/H0911" from BITS Pilani.

%

\normalsize\clearpage

\begin{onecolumngrid}
	\begin{center}
    
		{\fontsize{12}{12}\selectfont
        
			\textbf{Supplementary Material for ``Adiabatic charge transport through non-Bloch bands''\\[5mm]}}
            
		{\normalsize  Dharana Joshi,$^{1}$ Tanay Nag,$^{1}$\\
		{\small $^1$\textit{Department of Physics, BITS Pilani-Hyderabad Campus, Telangana 500078, India}\\[0.5mm]}}
		{}
	\end{center}
	
	\newcounter{defcounter}
	\setcounter{defcounter}{0}
	\setcounter{equation}{0}
	\renewcommand{\theequation}{S\arabic{equation}}
	\setcounter{figure}{0}
	\renewcommand{\thefigure}{S\arabic{figure}}
	\setcounter{page}{1}
	\pagenumbering{roman}
	
	\renewcommand{\thesection}{S\arabic{section}}
	
	

\section{Open-bulk Winding number using real space OBC }
\label{sm3}


In the main text, we demonstrate the real space winding number under OBC as shown in Fig. 2.  In this section, we discuss the formalism to compute the winding number for the open bulk case where the NH Hamiltonian is written in the real space as given below 
\cite{Song2019,He_2021,dharana2025}
\begin{equation}
    W=\frac{1}{2L}\mbox{Tr}'\Big(\tilde{\Gamma}Q[Q,\tilde{X}]\Big).
\label{rwind}
\end{equation}
Here, $\tilde{X}$ denotes the sub-lattice extended position matrix of dimension $(Ld\times Ld)$. $\tilde{X}=X_L\otimes I_d$ and $(X_L)_{mn}=m\delta_{mn}$, and   $I_d$ identifies the identity matrix of size ($d\times d$), $d$ refers to the sub-lattice degrees of freedom in that unit cell. $\tilde{\Gamma}$ designates the lattice extended chiral symmetry operator with  $\tilde{\Gamma}=I_L\otimes\Gamma$ where $I_L$ being the $L\times L$ identity matrix and $\Gamma$ is the generator of the chiral symmetry, represented by a $d\times d$ unitary matrix.  $Q$ denotes a ($Ld \times Ld$) matrix which is unitary as well as Hermitian,  satisfying $Q^2=I$.  $Q=\sum_{E_n>0}\ket{u_n^R}\bra{u_n^L}-\sum_{E_n<0}\ket{u_n^R}\bra{u_n^L}
$ with  $ H\ket{u_n^R}=E_n\ket{u_n^R},H^{\dagger}\ket{u_n^L}=E^*_n\ket{u_n^L}$. 
This is referred to as flattend Hamiltonian due to binary nature of eigenvalues. Note that Tr$'$ denotes the partial trace where we restrict the summation to the middle segment of the system $L'$ excluding $2l$ number of boundary sites from both the ends such that $L' + 2 l =L$. In this way, we obtain  quantized results of open-bulk winding number $W$ when $l$ is sufficiently large to avoid boundary effects. We  derive the above quantities from the real space OBC NH Hamiltonian $H=\sum_{n=1}^{L}\big[
(a+\gamma)A_nB^{\dagger}_n + (a-\gamma)B_n A^{\dagger}_n \big] + \sum_{n=1}^{L-1} \big[ b A_{n+1}B^{\dagger}_{n}+ b B_{n}A^{\dagger}_{n+1}+ c A^{\dagger}_{n}B_{n+1} + c B^{\dagger}_{n+1} A_{n}]+ \sum_{n=1}^{L-2} \big[
 d A^{\dagger}_{n+2}B_{n}+ d B^{\dagger}_{n}A_{n+2}$ \big]. Importantly, we do not renormalize the hopping as the non-Bloch momentum does not appear under OBC. The calculation of the OBC winding number does not depend on any specific value of $\beta$, since it is based on a finite system with  open boundary rather than an infinite system with bulk periodicity.


\section{Profile of complex energy spectrum over the phase diagram}
\label{sm7}

\begin{figure}
    \centering
    \includegraphics[width=1\linewidth]{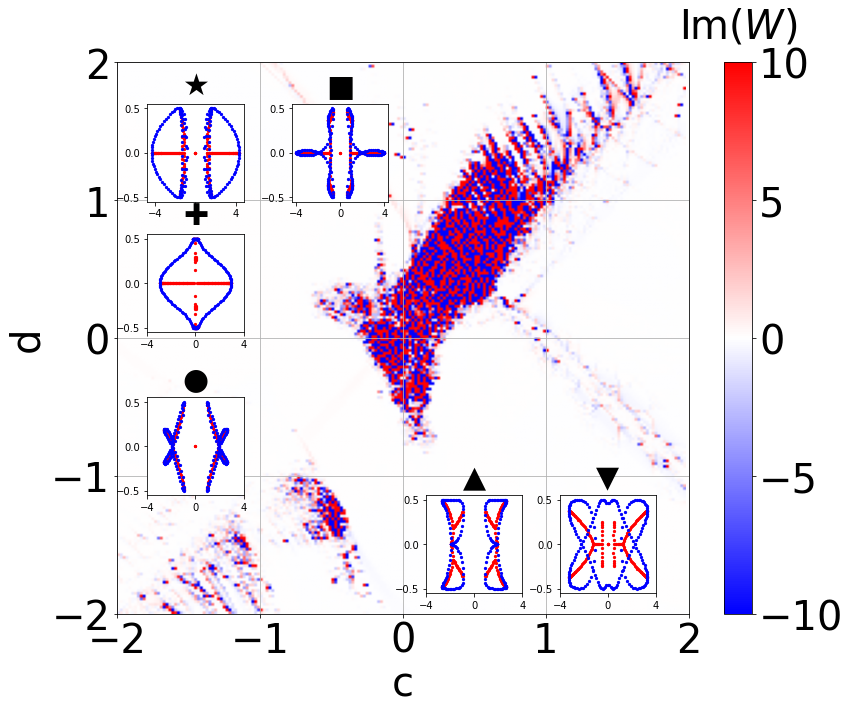}
    \caption{The main figure [inset] represents the imaginary part of the open-bulk winding number ${\rm Im}[W]$ in $c$-$d$ plane [${\rm Re}(E)$ vs ${\rm Im}(E)$ corresponding to their symbol taken from Fig. 2(a) of the main text.}
    \label{fig:obcpbc}
\end{figure}

In the main text, we discussed the real and imaginary energies of the NH model. It would be interesting to study the real and complex spectra over various spots on the phase diagram shown in Fig. 2(a) of the main text.   Figure \ref{fig:obcpbc} represents the imaginary part of the open-bulk winding number ${\rm Im}[W]$ that identifies the gapless region nicely. The inset in Fig. \ref{fig:obcpbc} demonstrates ${\rm Re}(E)$ vs ${\rm Im}(E)$ to study the energy vorticity in the complex energy plane under OBC (red dots) and PBC (blue dots) over different points on the phase diagram such that zero-energy modes are distinguished.  Note that $E$ is obtained from from the eigenvalues of the real space NH Hamiltonian where PBC Hamiltonian contains the additional terms $(b A_{1}B^{\dagger}_{L} + d A^{\dagger}_{2}B_{L} +{\rm H.C} )$ with respect to the OBC Hamiltonian.
The different insets indicate separate points in the phase diagram spanned by $c$ and $d$ and symbols are taken from Fig. 2(a) of the main text. 
This allows us to study gapped and gapless phases separately.  Remarkably, the PBC spectrum completely encloses the OBC spectrum while there exists a line gap in PBC for all gapped phases irrespective of topology, see $\star$, $\square$, $\triangle$, $\circ$,  $\triangledown$ marked figures. The point gap structure of PBC is observed in the critical NH phase as marked by $+$ symbol.  We find that OBC spectrum does not follow PBC spectrum at all for the critical phases while for off-critical phases they partially overlap with each other. The NH critical phase contains macroscopic number of states on ${\rm Re}(E)=0$ line.  The topological (trivial) gapped phases host (does not host) zero-energy modes residing at the origin of the complex plane for OBC, see $\triangle$ and $\circ$ symbols for example.


\section{Roots of characteristic equation: analytical form of complex momentum}
\label{sm1}


In this section, we analytically explore the non-Bloch form of the momentum from $H(\beta)$ with $e^{\pm ik} \to (\beta)^{\pm 1}$ which is given in Eq. (1) of the main text.  The Hamiltonian is given by 
\begin{equation}
H(\beta)=\begin{bmatrix}
  0 & z_1  \\
  z_2 & 0 
  \end{bmatrix} =
\begin{bmatrix}
 0 & (a+\gamma)+ b\beta^{-1}+c\beta +d\beta^{-2} \\
 (a-\gamma) +b\beta+c\beta^{-1} +d \beta^{2} & 0    
  \end{bmatrix}
\label{eq:sshlr2}  
\end{equation}
Here $\beta$ is a complex number and $z_1^* \neq z_2$  representing the NH case. The characteristic equation obtained from the characteristic equation ${\rm det}[H(\beta)-E]=0$ which gives $f(\beta) + E^2=0$ with  
\begin{eqnarray}
f(\beta)&=&
-\beta^6 cd-\beta^5\gamma d -\beta^5ad-\beta^5 bc-\beta^4 \gamma b+\beta^4 \gamma c-\beta^4 ab-\beta^4 ac 
-\beta^4 bd +\beta^3 \gamma^2-\beta^3 a^2-\beta^3 b^2-\beta^3 c^2 \nonumber \\
&-&\beta^3 d^2+\beta^2 \gamma b - \beta^2 \gamma c 
-\beta^2 ab -  \beta^2 ac-\beta^2 bd+\beta \gamma d-\beta ad-\beta bd -cd+\gamma^2
\label{eq:Eneq}
\end{eqnarray}
This is a sixth degree ploynomial equation in $\beta$, therefore, we get six roots  of $\beta$. Given the fact that $\beta$ is a complex number, we can use the gauge freedom and exercise the following equality  $f(\beta)=f(\beta e^{i\theta})$ leading to the equation as given below
\begin{eqnarray}
    &&\beta^3cd\big(e^{3i\theta}-1\big)+\frac{1}{\beta^3}cd\big(e^{-3i\theta}-1\big) + \beta^2\big(-\gamma d-ad-bc\big)\big(1-e^{2i\theta}\big) +\frac{1}{\beta^2}\big(\gamma d-ad-bc\big)\big(1-e^{-2i\theta}\big) \nonumber \\
    &+&\beta \big(\gamma c-\gamma b-ab-ac-bd\big)\big(1-e^{i\theta}\big)+\frac{1}{\beta}\big(\gamma b-\gamma c-ab-ac-bd\big)\big(1-e^{-i\theta}\big)=0
    \label{eq:chareq}
\end{eqnarray}
To determine the exact value of $\beta$, we explore an analytical approach to solve the equation. Solving Eq. (\ref{eq:chareq}) would lead to the formation of aGBZ when $\theta$ is varied from $0$ to $2\pi$, as shown in Fig. \ref{fig:aGBZ}. Note that the numerical 
method for computing the  non-Bloch momentum exactly is already described in the main text.

We have a gauge freedom here where two solutions say  $\beta$ and $\beta'$ have same absolute value i.e. $|\beta|$=$|\beta'|$ but  $\beta' =\beta e^{i \theta }$. Therefore, one can obtain three pairs of analytical solutions.  
We simplify the Eq. (\ref{eq:chareq}) by grouping terms with identical positive and negative powers of $\beta$ such as $\beta^n$ and $\beta^{-n}$ with $n=1,2$ and $3$. This allows us to assemble terms with $(e^{\pm ni\theta}-1)$ and separately equate them to zero. The individual group sum is zero ensuring  the total sum to be zero as well.  From the first group with $\beta^{\pm 3}$, we obtain the following
\begin{equation}
    \beta^6 =\frac{1-e^{-3i\theta}}{e^{3i\theta-1}}
    \label{eq:beta6}
\end{equation}
The above Eq. (\ref{eq:beta6}) corresponds to six roots while each of them having  $|\beta|=1$, representing the Hermitian case. Therefore, this solution is not useful for the present NH case. We continue with $\beta^{\pm 4}$ term giving the following 
\begin{equation}
    \beta^4 =\frac{ad+bc-\gamma d}{ad+bc+\gamma d }\bigg(\frac{1-e^{-2i\theta}}{e^{2i\theta -1}}\bigg) 
  \label{eq:beta4}
\end{equation}
Therefore, Eq. (\ref{eq:beta4}) leads to four roots with 
\begin{equation}
    |\beta|=\bigg(\frac{ad+bc-\gamma d}{ad+bc+ \gamma d}\bigg)^{\frac{1}{4}}
\label{eq:beta41}
\end{equation}
In the similar fashion, $\beta^{\pm 2}$ term yields 
\begin{equation}
 \beta^2 =\frac{\gamma c-\gamma b+ab+ac+bd}{\gamma b-\gamma c +ab+ac+bd}\bigg(\frac{1-e^{-i\theta}}{e^{i\theta -1}}\bigg) 
\label{eq:beta2}
\end{equation}
with the absolute value 
\begin{equation}
    |\beta|= \bigg(\frac{\gamma c-\gamma b+ab+ac+bd}{\gamma b-\gamma c +ab+ac+bd}\bigg)^\frac{1}{2}.
\label{eq:beta21}
\end{equation}

For the sake of nomenclature, we refer to Eqs. (\ref{eq:beta21}) and (\ref{eq:beta41}) as $\beta_2$ and $\beta_4$ roots, respectively, while $\beta_6$ roots following Eq. (\ref{eq:beta6}) remain trivial. Here, the analytical expressions of $|\beta|$ for both cases are independent of the choice of $\theta$. 
It is important to note that the results obtained using $\beta_2$ root reproduce the $|\beta|$ correctly for nearest neighbour NH SSH model when long-range hopping terms are turned off by setting $c=d=0$. In this case, one can obtain two roots and $\beta_2$ would correspond to the correct roots with $|\beta|=\sqrt{(a-\gamma)/(a+\gamma)}$ \cite{Okuma_2023}. On the other hand, in the present context, due to the long-range nature of the model we have $\beta_4$ and $\beta_2$ roots, see Eqs. (\ref{eq:beta41}) and (\ref{eq:beta21}), that we will use to compute the topological phase diagram.


\section{Behavior of winding number using the analytical roots}
\label{sm2}


In the main text, we demonstrate the winding number following the numerical finding of the non-Bloch momentum. Note that exploiting the above roots of $\beta$ given in Eqs. (\ref{eq:beta41}) and (\ref{eq:beta21}), can be used to compute the winding number.   
We, therefore,  now describe the phase diagram obtained from the analytical roots. In the main text we mention that analytical roots can only explain the behavior of the system deep inside a   phase while the phase boundaries can not be explained by the analytical roots.  Exploiting the chiral symmetry, generated by $\Gamma=\sigma_3$,  $\tilde H= \Gamma_U H \Gamma_U^{\dagger}= H$ where $\Gamma_U$ represents the chiral basis, 
one can  compute the winding of $q$, obtained from the flattened Hamiltonian.
\begin{equation}
Q=\sum_{E_n(\theta)>0}\ket{u_n^R(\theta)}\bra{u_n^L(\theta)}-\sum_{E_n(\theta)<0}\ket{u_n^R(\theta)}\bra{u_n^L(\theta)}
\nonumber
\end{equation}
Note that  $ \tilde{H}\ket{u_n^R}=E_n\ket{u_n^R},\tilde{H}^{\dagger}\ket{u_n^L}=E^*_n\ket{u_n^L}$, here $\theta$ is generalized bloch momentum. Winding number for a fixed $|\beta|$,  being independent of $\theta$, is given by 
\begin{eqnarray}
W(\beta)
&=&
\frac{1}{2\pi i} \int_0^{2\pi} d\theta \, \partial_\theta \log\big[\det(q)\big] 
\label{eq:winding2}
\end{eqnarray}
with 
\begin{equation}
Q = \begin{pmatrix} 
0 & q^\dagger \\ 
q & 0 
\end{pmatrix}. \ 
\nonumber \\
\end{equation}
We can also compute the winding number using the other block, $q^\dagger $. This will give the same value but with the opposite sign.

There is another method also for calculating winding number in momentum space as described in \cite{dharana2025}. 
We can transform the Hamiltonian given in Eq. (1) of the main text or  Eq. (\ref{eq:sshlr2}), in Chiral basis. The Hamiltonian given in Eq. (1) of the main text or  Eq. (\ref{eq:sshlr2}) remains invariant once it is written in the chiral basis $\tilde H= \Gamma_U H \Gamma_U^{\dagger}= H$. This allows us to directly use  Eq. (1) of the main text to compute the block winding number, for a given value of $|\beta|$ independent of $\theta$, as given by 
\begin{equation}
    W_{i}(\beta)=\frac{1}{2\pi i }\int_0^{2\pi} d\theta z_i^{-1} \frac{dz_i}{d\theta} 
    \label{eq:winding_block}
\end{equation}
where $z_i$ is the $i$-th  off-diagonal block given in Eq. (\ref{eq:sshlr2}) and $\beta$ is replaced by $|\beta| e^{i\theta}$. 
Here, $\theta$ is analogous to the Bloch momentum. In other words, $\theta$ can be thought of as the periodic part of the non-Bloch momentum.
In general, for any Hamiltonian, the winding numbers observed from both the off diagonal blocks are typically equal and opposite. This holds true even in the case of the NH 4 band SSH  model. However, this is not true for the NH SSHLR model due to its long-range nature. In this model, the winding numbers obtained from the two blocks are different from each other.
The total winding number in the given topological region is given by 
\begin{equation}
    W(\beta)=\frac{W_2(\beta) -W_1(\beta)}{2}
    \label{eq:winddif}
\end{equation}
where $W_1(W_2)$ is the block winding number obtained from the bottom left (top right) off-diagonal block   \cite{verma2024non,verma2024topological,Yokomizo2019}.  One important correlation between $W_1$ and $W_2$ phase diagrams is that $W_{1,2}$ phases almost overlaps with $W_{2,1}$ when $d \to c$ and $c \to d$. This is commonly observed for  $\beta_4$ as well as $\beta_2$ cases. This can be understood from  the fact that $c\beta + d\beta^{-2}$ in $z_1$ and $c\beta^{-1} + d\beta^{2}$ in $z_2$ are related to each other  by a factor of $\beta$ and an exchange between $d$ and $c$. More precisely,  
$c\beta^{-1} + d\beta^{2} = \beta(c \beta^{-2}+d\beta)$ and when $c (d)$ is replaced by $d (c)$, the right hand side expression except the common multiplication factor $\beta$  matches exactly with 
$d \beta^{-2}+c\beta$ which is there in the $z_1$ of  Eq. (\ref{eq:sshlr2}). On the hand, due to factor $\beta$ this $c(d)$ to $d(c)$ conversion is not exact leading to two slightly different phase diagrams for $W_{1,2}$ even when the axes are interchanged.

\subsection{$\beta_2$ root}

To obtain the topological phase diagram using the root with modulus value $|\beta_2|$, given in Eq. (\ref{eq:beta21}), we substitute the value of  $\beta= |\beta_2| e^{i\theta} $ in Eq. (\ref{eq:sshlr2}).
The topological phase diagram, obtained from Eq. (\ref{eq:winding2}), is shown in Fig \ref{fig:Supp3}(a), where we obtain integer quantized values $W(\beta_2)=2$ and $-1$, depicted by red and blue  regions, respectively. We observe half-integer values as well, and this half-integer region is situated in the region where we observe fluctuations in the OBC winding number.  To be precise, $W(\beta_2)=2$ and $-1$ phases are prominently observed while $W(\beta_2)=1,0$ phases are marginalized in $d,c \to 0$ window as compared to the Hermitian phases. At the same time, $W(\beta_2)= 1/2$ and $3/2$ phases with significant extents appear close to the phase boundaries, along with the fact that  $W(\beta_2)=1,0$ phases emerge between $W(\beta_2)=2$ and $-1$ phases.

\begin{figure}[ht]
    \centering
    \includegraphics[width=1\linewidth]{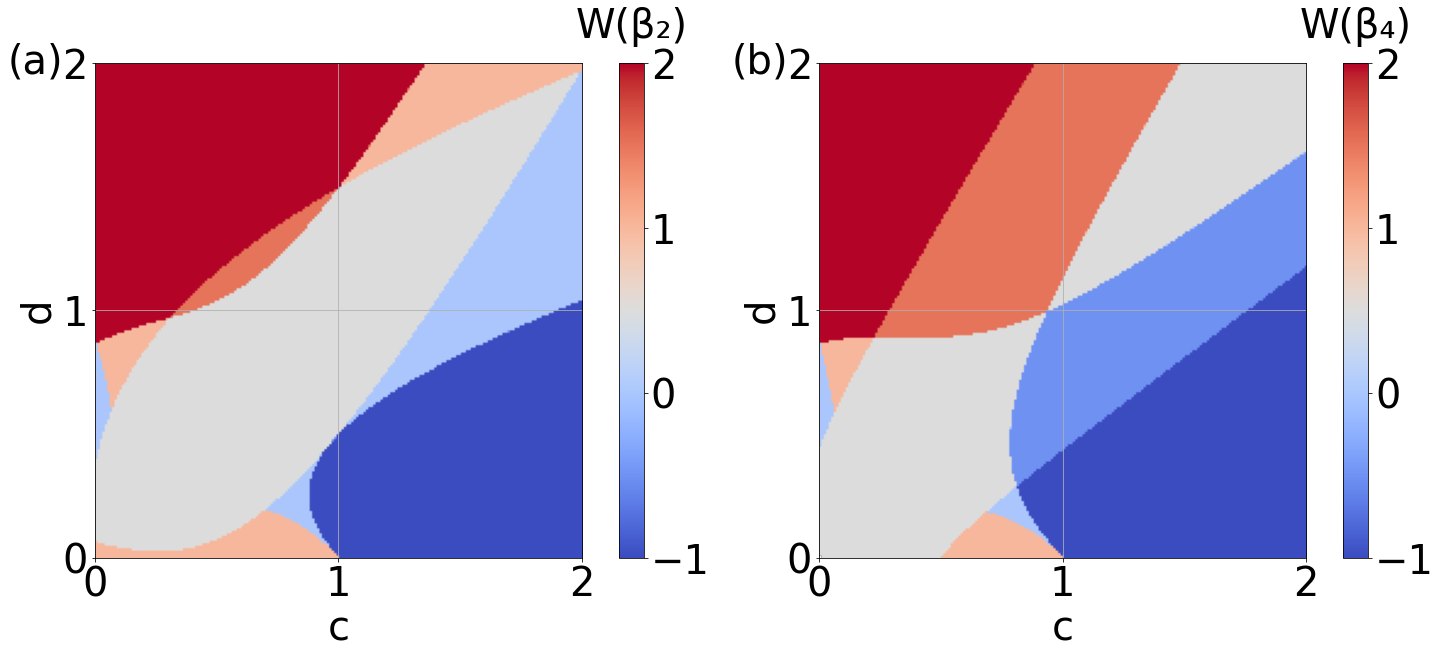}
    \caption{Topological phase diagram following analytical root $|\beta_2|$ Eq.  (\ref{eq:beta21}) and   $|\beta_4|$ Eq. (\ref{eq:beta21}) in (a), and (b), respectively. Here, $a=b=1, \gamma=0.5$, number of momentum modes i.e., number of discretized $\theta$ is $N=200$, however, $|\beta|$ does not depend on $\theta$. The colorbar represents winding number Eq. (\ref{eq:winddif}).}
    \label{fig:Supp3}
\end{figure}


Now coming to the block winding number $W_{1,2}(\beta_2)$, we find the following.  
Interestingly, two blocks of the Hamiltonian yield two different phase diagrams in NH SSHLR model which is in complete contrast to the NH  4 band SSH model where both the blocks result in exactly opposite  phase diagrams.
The second-nearest neighbour hopping introduces new features causing $W_1 (\beta_2) \ne -W_2(\beta_2)$ across the parameter space.  Therefore, the phase boundaries are significantly dissimilar between $W_1(\beta_2)$ and $W_2(\beta_2)$ while deep inside the phase one can find $W_1(\beta_2) =-W_2(\beta_2)$.

\subsection{$\beta_4$ root}

Topological phase diagram using $\beta_4$  roots, given in Eq. (\ref{eq:beta41}), is obtained by replacing $\beta$ with 
$|\beta_4| e^{i\theta}$ as shown in Fig \ref{fig:Supp3} (b). Similar to the $\beta_2$ analysis,  we repeat the winding number computation with $\beta_4$. One can find similar features to those of $\beta_2$ roots. We obtain $W(\beta_4)=2,-1$ phases, but phase boundaries enclosing $W(\beta_4)=1,0$ regions are significantly modified. The middle portion of the plot, which shows the half-integer winding number, is now broaden as compared to Fig \ref{fig:Supp3} (a). Using both the analytical roots $\beta_2$ and $\beta_4$ phase diagram using winding number is not able to capture the correct phase boundaries.

Now comparing with the Hermitian phase diagram, NH clearly adds richness when we observe $W(\beta_4)=\pm 1/2$ and $3/2$ phases. Interestingly, there exists a contrast compared to  the $\beta_2$ phase diagram where $W(\beta_4)=-1/2$ phase was absent. The $\beta_4$ phase diagram is more deviated from the Hermitian phase diagram as the $W(\beta_4)=1,0$ phases almost become non-existent there in the former case within the window  $d,c>0$. Since this analytical approach is not able to accurately predict the phase boundaries for a complex model like NH SSHLR, a numerical approach is necessary, flowchart for the procedure of calculating GBZ and $|\beta|$ is provided in detail in the following section.

\begin{figure}[ht]
    \centering
    \includegraphics[width=1\linewidth]{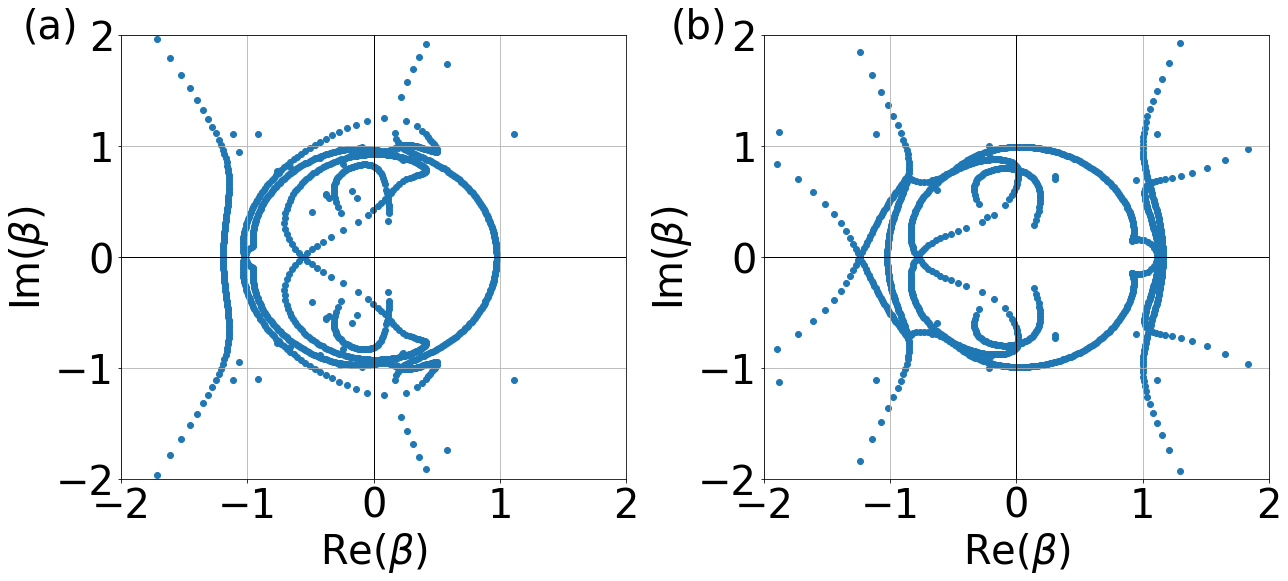}
    \caption{
    We show the aGBZ by solving Eq. (\ref{eq:chareq}) with $c=-1.5$ and $1.5$ in (a) and (b), respectively. We consider   $a=b=d=1,\gamma =0.5$, and number of discretized $\theta$ is $N=500$.}
    \label{fig:aGBZ}
\end{figure}


\section{Auxiliary generalized Brillouin zone (aGBZ) to generalized Brillouin zone (GBZ)}\label{sm4}

\begin{figure}
    \centering
    \includegraphics[width=0.7\linewidth]{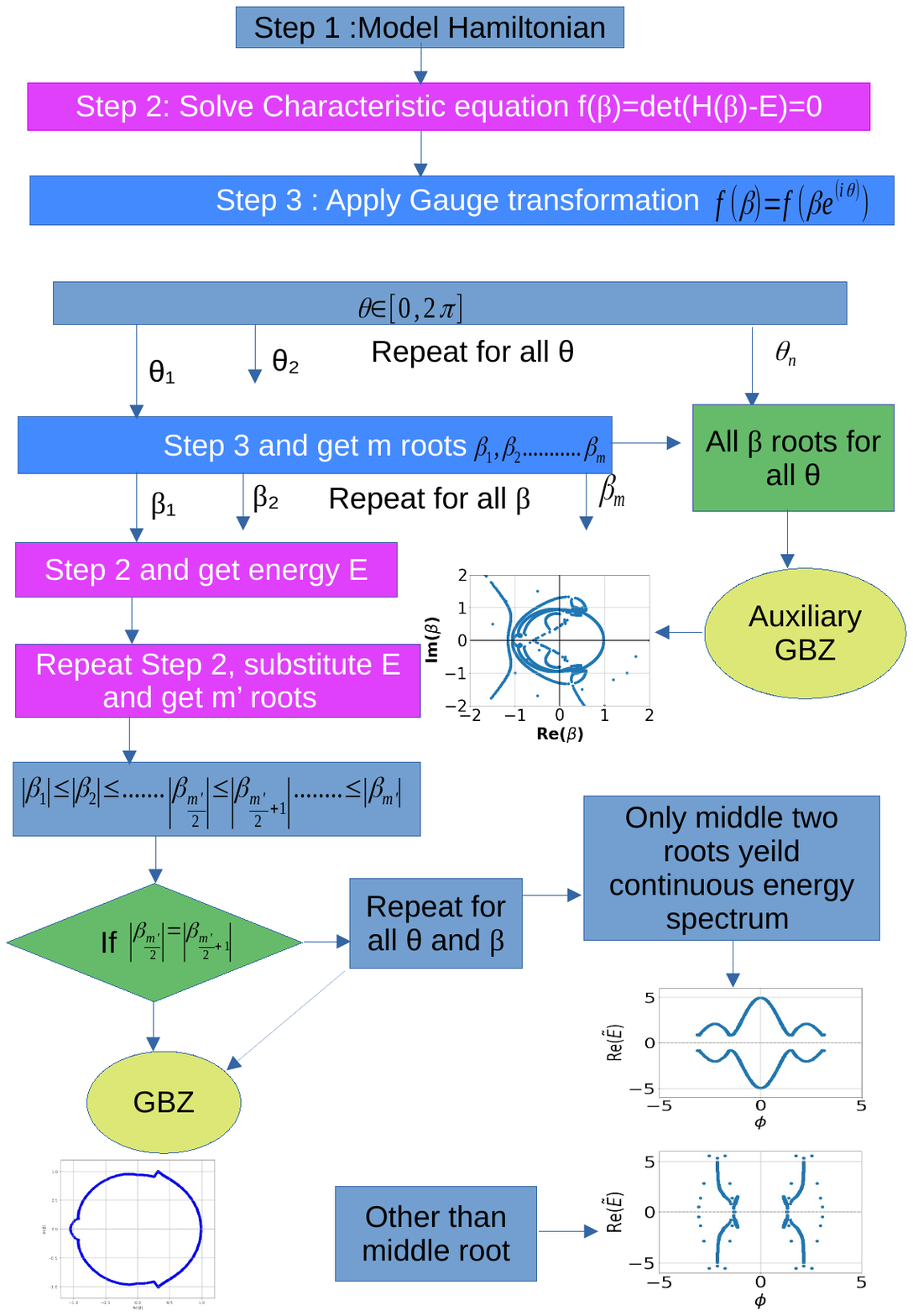}
    \caption{Flowchart for extraction of GBZ from the aGBZ. The steps are color-coded to highlight their iterative nature.    }
    \label{fig:flowchart}
\end{figure}

In the main text, we describe the process to obtain the GBZ from aGBZ. We note that the aGBZ is obtained by solving Eq. (\ref{eq:chareq}) and energy equation $f(\beta)+E^2=0$ for different values of $\theta\in [0, 2 \pi]$. To be precise, we have two equations to solve and two unknowns to obtain, $|\beta|$ and $E$ for a given $\theta$.  We here discuss in the detail the formation of GBZ from aGBZ. 
As we describe in the main text, by plotting the primary roots $\beta_p$ of Eq. (\ref{eq:chareq}) as $\theta \in (0,2\pi)$, we can plot the real versus imaginary part of the primary root giving the aGBZ, which is shown in Fig. \ref{fig:aGBZ}. Here we plot aGBZ in (a)[b] for parameter $c=-1.5[1.5]$ while keeping other parameter fixed $a=b=d=1,\gamma=0.5 $. From the figure, we can clearly see that there exists a large number of multi-valued $\beta$ roots, and it is difficult to identify the accurate path of  $\beta$ from the aGBZ. We present a clear and systematic procedure for extracting the GBZ from the aGBZ, illustrated in the form of a flowchart as shown in Fig. \ref{fig:flowchart}. The flowchart outlines the step-by-step construction of the aGBZ and then the extraction of GBZ from aGBZ using the characteristic equation by analyzing the complex $\beta$ roots self-consistently.

The primary roots obtained from Eq. (\ref{eq:chareq}) are first substituted into Eq. (\ref{eq:Eneq}) to compute the energy $E$ associated with each primary root $\beta_p$. These energy $E$ is then substituted back into Eq. (\ref{eq:Eneq})  to obtain the corresponding secondary roots $\beta_s$. This procedure ensures consistency between $\beta$ and the energy $\tilde{E}$ for a given value of $\theta$.
The secondary roots are then arranged in ascending order of their absolute values $  | \beta_{s1} | \le |\beta_{s2}|.........|\beta_{s5}| \le|\beta_{s6}|$. 
The GBZ is defined by the trajectory of only those secondary roots over the complete $\theta $ profile where $\theta \in (0, 2 \pi )$ for which the middle two roots have same absolute value such as $|\beta_{s3}|= |\beta_{s4}|$. Although other pairs of roots may also have equal absolute values(e.g., $|\beta_{s1}|=|\beta_{s2}|$), but only the condition on the middle pair yields the continuous energy bands ${\tilde E}$ associated with the bulk spectrum\cite{Yokomizo2019}. Interestingly, in cases where $c$ or $d$ is small but nonzero (e.g.,
$c,d \approx 0$), numerical solvers may fail to return all six roots due to near-degeneracy, occasionally yielding only five roots. In such scenarios, we again retain the second and third roots by modulus to ensure consistent selection of the physically meaningful modes. This root selection procedure is essential for maintaining continuity and correctness in the GBZ construction across parameter regimes.  
Using the resulting GBZ, the non-Bloch energy spectrum $\tilde E$ is also computed which is found to be continuous. Along with the flowchart in Fig. \ref{fig:flowchart}, we also provide  representative 
figures displaying aGBZ, GBZ, the continuous and discontinuous non-Bloch energy spectrum. The continuous spectra ${\tilde E}$ is  obtained by selecting the middle two secondary roots with equal absolute values, and the discontinuous energy spectrum is obtained when we  choose roots other than the middle ones.

Once the GBZ is obtained, the topological feature can be captured by plotting the quantities $d^{x,y}_{R}(\beta)$ when the polar angle $\phi={\rm arctan}\big[{\rm Im[\beta]}/{\rm Re[\beta]}\big]$ is varied between $-\pi$ to $\pi$. The inclusion of the origin there is an essential feature of the Hermitian topology. In the present NH case, the inclusion of any one out of two exceptional points $[d^x_R(\beta),d^y_R(\beta)]=[\mp d^y_I, \pm d^x_I]=[\mp \gamma,0]$ determines the topology \cite{Yin18}. The exceptional points satisfy both the conditions $d^2_R=d^2_I$ and ${\bm d}_R.{\bm d}_I=0$. The non-Bloch winding number $W_{\beta}=n$ of a topological gapped phase 
is caused by the inclusion of at least one of the exceptional points $n$ number of times in the parametric plane. The trivial gapped phase is associated with no enclosure of an exceptional point. 
The critical phase is associated with the touching of the exceptional points in the parametric plane. This is what is seen in Fig. 2 (c) of the main text further confirming the emergence of a $\phi$ dependent $|\beta|$.

\newpage

\section{Open-bulk spatio-temporal Bott index under open boundary condition in real space}
\label{sm5}


We now compute the open bulk Bott index on a real-space  lattice under OBC with the time-dependent Hamiltonian  $H(t)=  (h_0/2i) A_{n,m} A^{\dagger}_{n,m+1} -(h_0/2i) A_{n,m+1} A^{\dagger}_{n,m} 
-  (h_0/2i)  B_{n,m} B^{\dagger}_{n,m+1} + (h_0/2i)  B_{n,m+1} B^{\dagger}_{n,m} 
+
(a+\gamma)A_{n,m}B^{\dagger}_{n,m} + (a-\gamma)B_{n,m} A^{\dagger}_{n,m} + b A_{n+1,m}B^{\dagger}_{n,m}+ b B_{n,m}A^{\dagger}_{n+1,m}+ t_0 A^{\dagger}_{n,m}B_{n+1,m} + t_0 B^{\dagger}_{n+1,m} A_{n,m} - (d_0/2) A^{\dagger}_{n,m}B_{n+1,m+1}  - (d_0/2) A^{\dagger}_{n,m}B_{n+1,m-1} 
-
(d_0/2) B^{\dagger}_{n+1,m+1} A_{n,m} -
(d_0/2) B^{\dagger}_{n+1,m-1} A_{n,m}
+ t_0 A^{\dagger}_{n+2,m}B_{n,m}+ t_0 B^{\dagger}_{n,m}A_{n+2,m} 
+(d_0/2) A^{\dagger}_{n+2,m}B_{n,m+1}
+ (d_0/2) A^{\dagger}_{n+2,m}B_{n,m-1}
+ (d_0/2) B^{\dagger}_{n,m+1}A_{n+2,m}+ (d_0/2) B^{\dagger}_{n,m-1}A_{n+2,m}
$. Here, $(n,m)$ represents the unit cell index along $(x,y)$ direction on the lattice.   The open-bulk Bott index for a system having size  $L_x \times L_y$,  is defined as \cite{He_2021, Song2019},
\begin{equation}
    B=\frac{2\pi}{\tilde{L}_x \tilde{L}_y} {\rm Im}\big({\rm Tr'}[P \tilde{X} P,P \tilde{Y} P]\big)
    \label{eq:bott_obc1}
\end{equation}
where $P$ is the projector onto the occupied states. Here, $\tilde{X}$ and $\tilde{Y}$ denote the sub-lattice extended position matrix of dimension $(L_xL_yd\times L_xL_yd)$.
For the NH Hamiltonian, this projector is constructed using the bi-orthogonal set of left and right eigenstates $\langle\psi_i^L|$ and $|\psi_i^R\rangle$, respectively,  associated with the Hamiltonian $H(t)$, $ P = \sum_i^{N_{\rm occ}} |\psi_i^R\rangle\langle\psi_i^L|$.  ${\rm Tr'}$ is restricted to a central region of size $\tilde{L}_x \times \tilde{L}_y $where $(\tilde{L}_i = L_i - 2l_i) $ for $i=x,y$, $l_i$ number of unit cells are discarded from both the sides.  This restriction is used to minimize boundary effects in the calculation and $x$ and $y$ represent the position operator in momentum and time space.


\begin{figure}
    \centering
    \includegraphics[width=1\linewidth]{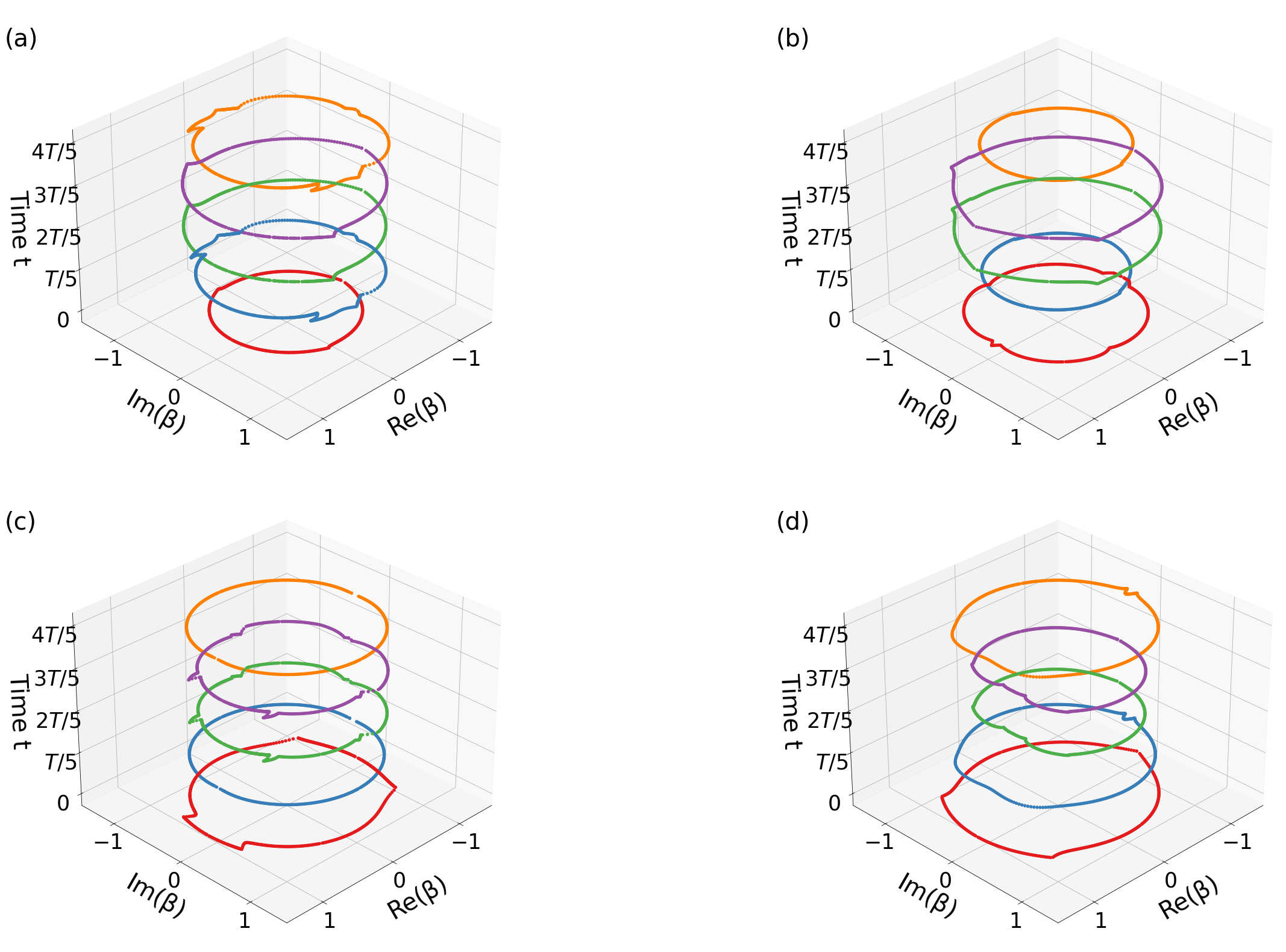}
    \caption{Variation of the GBZ over the time $t \in [0,T]$ where $x$ $(y)$ axis represents real (imaginary) part of the $\beta$ and time increases along  $z$ axis. We consider $(t_0,d_0)=(1.5,1.5), (0.5,1.5), (-0.5,1.5),(-1.5,1.5)$ in (a),(b),(c) and (d) respectively along with $h_0 =T= a= b=1, \gamma=0.5$ and number of momentum modes i.e., number of $\phi$ is  $N=200 $. We use $t=0, T/5, 2T/5, 3T/6, T$.}
    \label{fig:Supp5}
\end{figure}


\section{Time-dependent generalized Brillouin zone}
\label{sm7}

\begin{figure}
    \centering
    \includegraphics[width=0.9\linewidth]{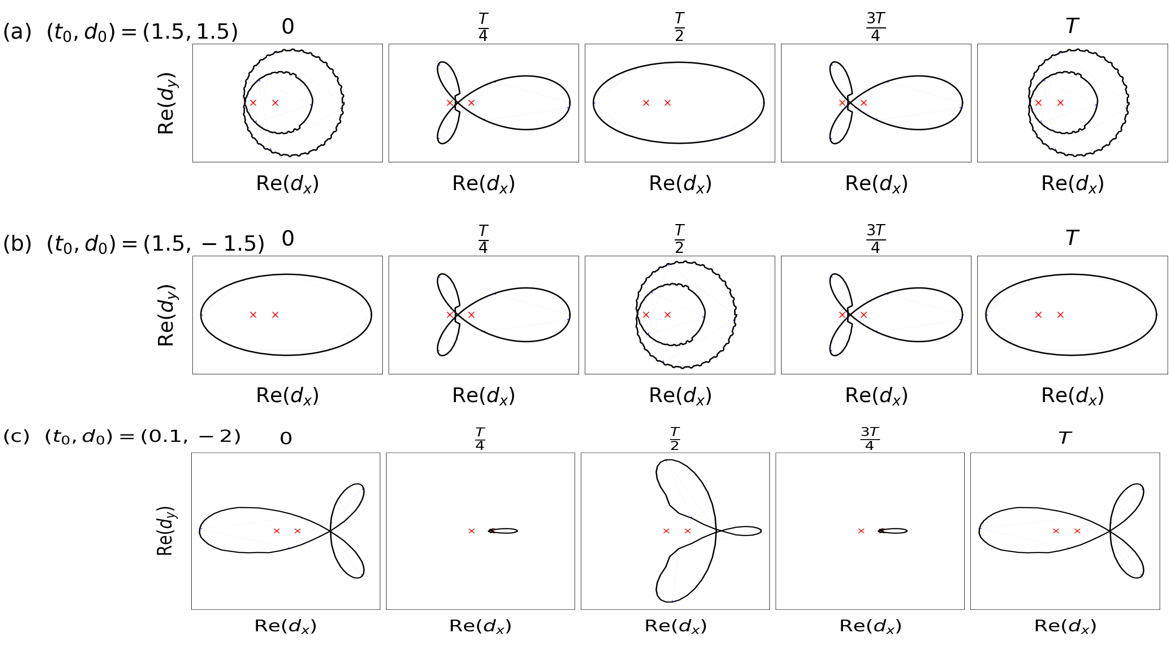}
    \caption{Parametric plot in $d^x_R(\beta,t)$-$d^y_R(\beta,t)$ plane for $t=0, T/4, T/2, 3T/4, T $ with  $(t_0,d_0)= (1.5,1.5), 
    (1.5,-1.5), (0.1,-2)$  in (a,b,c), respectively, indicating the Bott index $B=3$, $-3$, and $-1$ of the underlying phases.}
    \label{fig:windt}
\end{figure}

In this section, we show the evolution of the GBZ with time when we introduce time in the NH SSHLR model to study the adiabatic charge transport. This helps us to understand the  Fig. 4 (a) of the main text where we discussed the evolution of the non-Bloch energy bands with time. 
As time is introduced into the Hamiltonian, the aGBZ becomes explicitly time-dependent. Therefore, for a given time, one has to compute the GBZ from the aGBZ following the procedure discussed in Sec. \ref{sm4}. We repeat this calculation for all time $t\in [0,T]$ to obtain the evolution of GBZ with time.  We numerically obtain $|\beta|$ for each $\phi$ for all time $t \in [0,T]$ to study the evolution of GBZ with time. 
In Figs. \ref{fig:Supp5} (a,b,c,d),   we illustrate the evolution of the GBZ over time $t \in [0,T]$ for the sets of parameter  $(t_0, d_0 )= (1.5,1.5), (0.5,1), (-0.5,1)$ and $(-1.5,1.5)$  respectively. Among these,  Figs. (a), (c), and (d) correspond to topologically non-trivial phases, denoting off-critical phases, with Bott index  $B=-3,-1$ and $B=3$  respectively. In contrast, Figs. (b) represents a topological phase boundary which is critical point. As discussed in the main text, the critical and off-critical GBZ lead to gapless and gapped non-Bloch dispersion, see Fig. 4 (c1, c2) of the main text.  
We depict the GBZ at various time steps using different colors to visualize how it evolves as a function of time.  The closed nature of the GBZ with time clearly suggests that underlying NH non-Bloch bands are analogous to Hermitian Bloch bands.

Here we also show the parametric plot in $d^x_R(\beta,t)$-$d^y_R(\beta,t)$ plane for $t=0, T/4, T/2,3T/4,  T$ as shown in Fig. \ref{fig:windt}.
The topology of the GBZ can only be understood when $d^{x,y,z}_R(\beta,t)$ wind each other keeping the exceptional point $[d^{x}_R(\beta,t),d^{y}_R(\beta,t),d^{z}_R(\beta,t)]=[\mp \gamma, 0, 0]$ inside them. 
While comparing with the static situation, one can find that the inclusion of the exceptional point occurs with time for the adiabatically driven case. Importantly, $d^R_z(\beta,0)=d^R_z(\beta,T/2)=0$ leading to the inclusion of exceptional point in the $d^R_z(\beta,t)=0$-plane where the static understanding holds true. 
This feature with time is directly extended from the Hermitian case with Bloch momentum modified with non-Bloch momentum.  For example,  for Bott index $B = +3$, the trajectory winds twice at $t = 0$ and once at $t = T/2$ around the exceptional point $[d^{x}_R(\beta,t),d^{y}_R(\beta,t),d^{z}_R(\beta,t)]=[\mp \gamma, 0, 0]$, whereas for $B = -3$, it winds once at $t = 0$ and twice at $t = T/2$, see  Fig. \ref{fig:windt}(a,b) for $B=3$ and $-3$, respectively. The above is consistent with the fact that two zero energy modes at $t=0$ and one zero energy mode at $t=T/2$. On the other hand, $B=-1$ phase is also replicated by the single winding in  
$d^{x}_R(\beta,t=0)$-$d^{y}_R(\beta,t=0)$ plane around the exceptional point while there is no winding at $t=T/2$, see Fig. \ref{fig:windt}(c). The chirality of winding determines the sign of the Bott index. In the case of the critical phases,  the winding in 
$d^{x}_R(\beta,t)$-$d^{y}_R(\beta,t)$ plane  touches the exceptional point at $t=0$ and $T/2$.

In the Fukui formalism of computing the Chern number as discussed in the main text, it is important to have a consistent grid in both momentum and time to evaluate the Berry curvature accurately. However, our numerical results show that the GBZ at different times can get different values of the phase. This happens because the phases are obtained by back-calculating from the complex $\beta$ roots, which change with time. As a result, the time-dependent GBZ makes it challenging to keep a uniform phase grid i.e., the values of polar angle $\phi={\rm arctan}\big[{\rm Im[\beta]}/{\rm Re[\beta]}\big]$, for the Chern number calculation. At different time slices do not naturally align due to the evolving nature of the GBZ, we employ an interpolation scheme to reconstruct the GBZ data on a common set of polar angle points $\phi \in [0, 2\pi]$ across all times. Along with this, we interpolate the modulus of the complex parameter $|\beta|$ at those fixed polar angle points and substitute it into the Chern number formalism. This consistent handling of both the polar angle and magnitude ensures that the discretized grid remains uniform throughout the time evolution, allowing for accurate Chern number calculations and smooth tracking of topological features.  In the Fig. 4 (d1,d2,d3,d4) of the main text, we have followed the above procedure to compute the non-Bloch Chern number. 


\section{Quantized pumping obtained from analytical form of non-Bloch momentum}
\label{sm6}

In this section, we investigate the adiabatic evolution of a non-Hermitian system by treating time as an additional parameter in the Hamiltonian. While the numerical method for computing the Chern number has been thoroughly discussed in our main text and Fig. 4, here we complement that analysis by exploring an analytical approach to evaluate the Chern number. As discussed in the static case, the analytical result yields the correct topological invariant deep inside the phase but fails to accurately capture the phase boundary. A similar situation occurs for the time-dependent Hamiltonian, where the Chern number is correctly obtained deep inside the phases but not near the phase boundaries. While the numerical approach is extensively discussed in the main text, here we discuss the analytical approach.

In adiabatic charge transport, the time is treated as another variable along with momentum $k$, allowing us to define the 2D Chern number. Similar to non-Bloch winding number ,we can calculate the non-Bloch Chern number by replacing $H(k)$ with $H(|\beta|e^{i\theta})$, where $\theta $ is new generalized momentum. Note that $|\beta|$ is 
considered to be independent of $\theta$. 
In case of NH matrix $ H \neq H^\dagger$, the right and left eigenstates satisfy the
following eigenvalue equations $H(\theta)\ket{u_n^R(\theta)}=E_n\ket{u_n^R(\theta)},\quad H(\theta)^{\dagger}\ket{u_n^L(\theta)}=E^*_n\ket{u_n^L(\theta)}$.
The non-Bloch Chern number for the \(n\)th band can be expressed as \cite{fukui2005chern}
\begin{equation}
C_n({\beta}) = \frac{1}{2\pi i} \int_{0}^{2\pi} \int_{0}^{T} d\theta \, dt \, F_{12,n}(\theta,t),
\label{CheNumCon}
\end{equation}
where $F_{12,n}(\theta,t) = \partial_1 A_{2,n}(\theta,t) - \partial_2 A_{1,n}(\theta,t)$ ($1,2=\theta,t$) denotes the Berry curvature and $A_{\mu,n}(\theta,t) = \bra{n(\theta,t)} \partial_\mu \ket{n(\theta,t)}$ represents the Abelian Berry connection.

\begin{figure}
    \centering
    \includegraphics[width=1\linewidth]{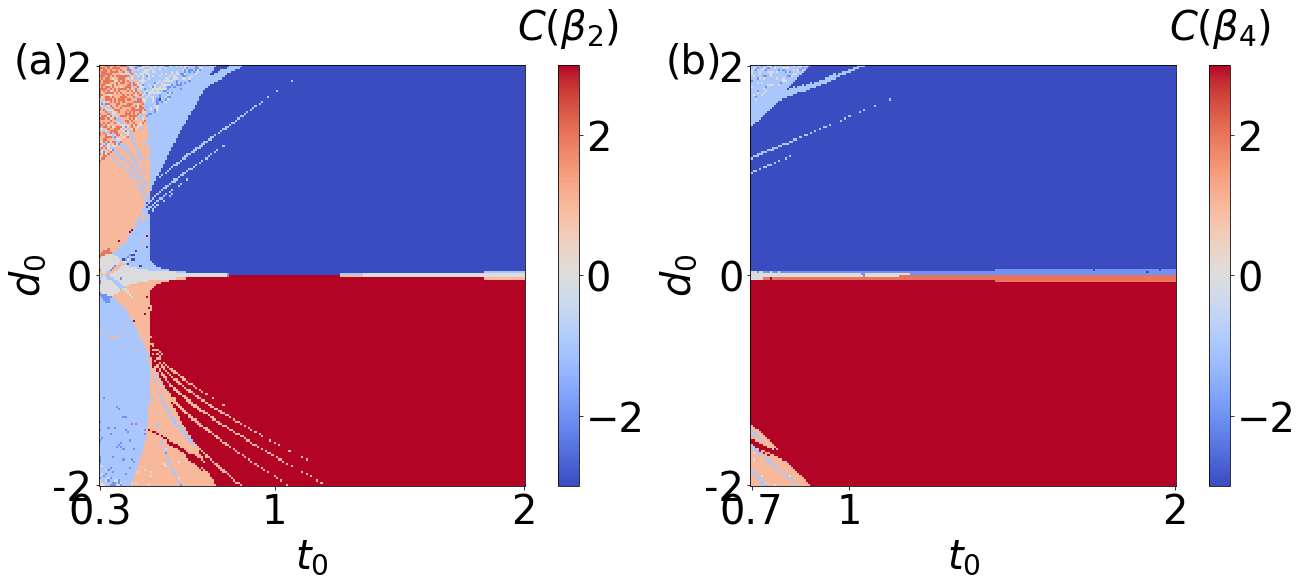}
    \caption{We show the topological phase diagram, obtained from non-Bloch Chern number, from $|\beta_2|$ Eq. (\ref{eq:beta21}) and $|\beta_4|$ Eq. (\ref{eq:beta41})
    in (a) and (b), respectively. Here number of momentum modes is $N=200$, $a=b=h_0=T=1$ and $\gamma=0.5$. }
    \label{fig:Chern}
\end{figure}



\begin{figure}
    \centering
    \includegraphics[width=1\linewidth]{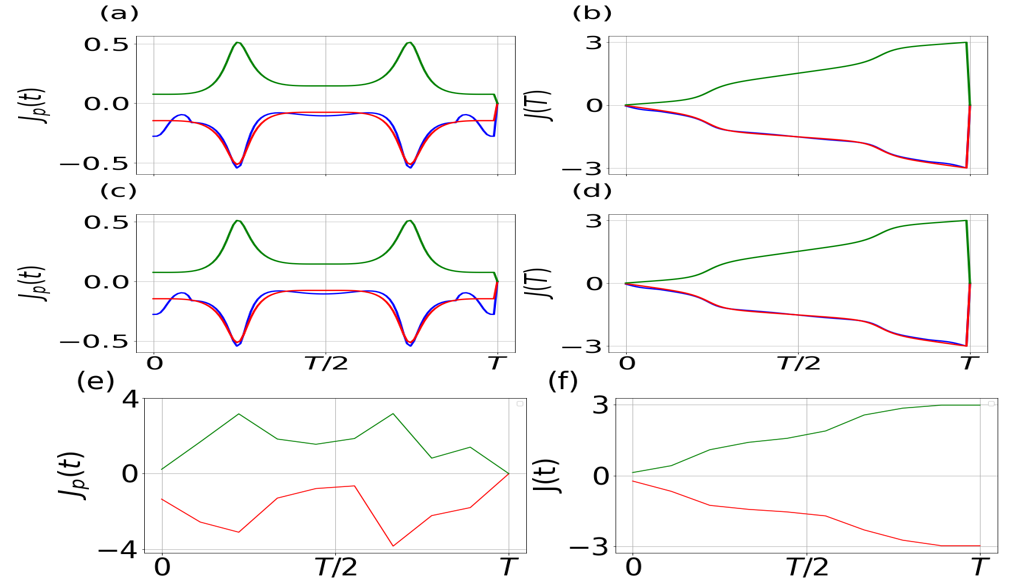}
    \caption{The time evolution of particle current $J_p(t)$ and pumped
charge $J(t)$  for analytical root $|\beta_2|$ in (a) and (b), respectively. We repeat (a,b) in (c,d) for analytical root $|\beta_4|$. We repeat (a,b) for the numerical root $|\beta|$ in (e,f). We choose $(t_0,d_0)$ = $(1.5, 1.5), (1.5, -1.5)$, represented by red, and blue, respectively. }
    \label{fig:current}
\end{figure}


We show the Chern number $C({\beta_2})$ and $C({\beta_4})$ for ground state with $n=1$, using the non-Bloch formalism using both $\beta_2$ Eq. (\ref{eq:beta21}) and $\beta_4$ Eq. (\ref{eq:beta41}),  in Fig \ref{fig:Chern} (a) and (b), respectively.  We consider $a = b = 1$ and $\gamma = 0.5$ while plotting the Chern number in the $t_0$-$d_0$ parameter plane. In Fig. \ref{fig:Chern} (a) with root $|\beta_2|$, we find Chern number $C({\beta_2})$ yields finite numbers for $0.3<t_0<2$.  On the other hand, 
$|\beta_4|$ also gives a similar phase diagram for $t_0>0.7$ only as Chern number $C({\beta_4})$  becomes undefined for $t_0 < 0.7$ as shown in Fig. \ref{fig:Chern}(b).
Therefore, we are limited to analyze the regime deep inside the boundary, where the system exhibits behavior similar to the Hermitian case.
For example, at $(t_0,d_0) = (1.5,\pm 1)$, deep within the phase, the Chern number is $C({\beta_2}), C({\beta_4})= \mp 3$.
However, as we approach the boundary line around $t_0 = 0.5$,  intermediate phases with $C({\beta_2}),C({\beta_4}) = \pm 1$ emerge along the axis of $d_0$. These phases do not show up in the Hermitian model.
In the main text, we compute the complete phase diagram using the real-space Bott index under OBC where the fluctuation of the invariant around the Hermitian phase boundaries is observed. From the analytical solution, we obtain intriguing behavior around the Hermitian phase boundaries. Therefore, the analytical approach is able to give a hint about the phase boundary while deep inside a phase analytical roots $|\beta_{2,4}|$ can predict the behavior of NH model nicely.  Importantly, the  phase diagram in the full parameter space can not be obtained by using  the analytically obtained roots while the numerical solution of characteristic equation captures the topology of model irrespective of the parameter region. 


We compute the 
particle current $J_p(t)=\sum_{\rm n\in occ}\frac{1}{2\pi i} \int_{0}^{2\pi}  dk  \, F_{12,n}(k,t)$ and pumped charge $J(t)= \int_{0}^{t}  dt' J_p(t')$ using the analytical roots $|\beta_2|$ as given in Eq. (\ref{eq:beta21}). One can repeat the calculations using root $|\beta_4|$ given in Eq.   (\ref{eq:beta41}). For deep inside a topological phase considering $|\beta_2|$, we show the time variation of the particle current $J_p(t)$ in Fig.\ref{fig:current}(a) while the pumped charge $J(t)$ is depicted in  Fig. \ref{fig:current}(b). We compare our findings for $|\beta_2|$ with $|\beta_4|$ roots as well in  Figs.\ref{fig:current}(c,d). $J_p(t)$ and $J(t)$ for $|\beta_2|$ and  $|\beta_4|$ are exactly the same indicating the fact that 
analytical roots $\beta_2$ and $\beta_4$  are able to characterize the topology for parameter points deep inside the phase.  Going beyond the analytical solution, we demonstrate the $J_p(t)$ and $J(t)$, obtained for GBZ where for each $\phi$ results in different values of $|\beta|$, see Figs. \ref{fig:current} (e,f). The numerical roots of $\beta$ result in qualitatively similar particle current and pumped charge as compared to their analytical counterparts.

As shown in Figs. \ref{fig:current} (a-f), the time profile of $J_p$ and $J$ vary quantitatively, however, qualitatively they share common features which are similar to their Hermitian counterparts. The even nature of $J_p(t)$ with respect to $t=T/2$ suggests that the full integral over time would be non-zero unless the area under the positive $J_p(t)$  curve cancels the area under the negative $J_p(t)$ curve. Interestingly, the $J(t)$ represents the amount of  the electrical charge passed from one edge to the other edge till time $t=t$. Therefore, the ups and downs of $J_p(t)$ till that instant $t=t$ will be reflected in the profile of $J(t)$. The pumped charge $J(t=T)$ at final time $t=T$ reaches an integer value as shown in \ref{fig:current} (b,d,f) referring to the  quantized charge transport.







\section{Time evolution of energy levels under OBC- a signature of BBC}
\label{sm7}

In this section, we  demonstrate the variation of bands with time while the real part of the energy under OBC is illustrated, see  Fig. (\ref{fig:pump}).
In the main text, we directly show the behavior of the spatio-temporal Bott index in Figs. 4 (a,c) of the main text. The analysis here helps us to understand that results  better from the flow of electrons with time.  
To further verify our findings on topological, critical phases in the driven case and BBC, we analyze the energy-time profile. 
In the topological phase, one can clearly find electron traveling from  one end (yellow  color denotes the left edge) to the other end (blue color denotes the right edge) of the chain when time $t$ changes from $0$ to $T$. One can identify the Chern number from the chirality of the crossing i.e., whether yellow goes up (down) and blue goes down (up) when the crossing takes place at $t=0$, $T/2$, and $T$.   The connection between  chiral crossing of the boundary modes having real energy with time and the bulk non-Bloch Chern number directly signifies the BBC and is analogous to the Hermitian counter part where  Chern number   of the bulk Bloch bands mimics the temporal behavior of the mid-gap boundary 
 states. 
In Figs.(\ref{fig:pump}) (a,b), we consider 
$(t_0, d_0) = (1, 2)$ and $(t_0, d_0) = (1, -2)$, respectively, to show the chiral profile of the crossings leading to the non-Bloch  Chern number $C = -3$ and $+3$.
The opposite nature of chirality 
around the crossing is clearly visible in 
Figs.(\ref{fig:pump}) (a,b).
Deviating from a point deep inside the topological phase, we find gapless behavior with macroscopic degeneracies in 
Fig.(\ref{fig:pump}) (c) for $(t_0, d_0) = (0.6, 2)$ which is inside an extended critical phases lying between two distinct topological phases.  Finally, in Fig.(\ref{fig:pump})(d), for $(t_0, d_0) = (0.3, 2)$, we observe the topological phase with $C = +1$ showing only one chiral crossing at $t=T/2$.

\begin{figure}
    \centering
    \includegraphics[width=0.9\linewidth]{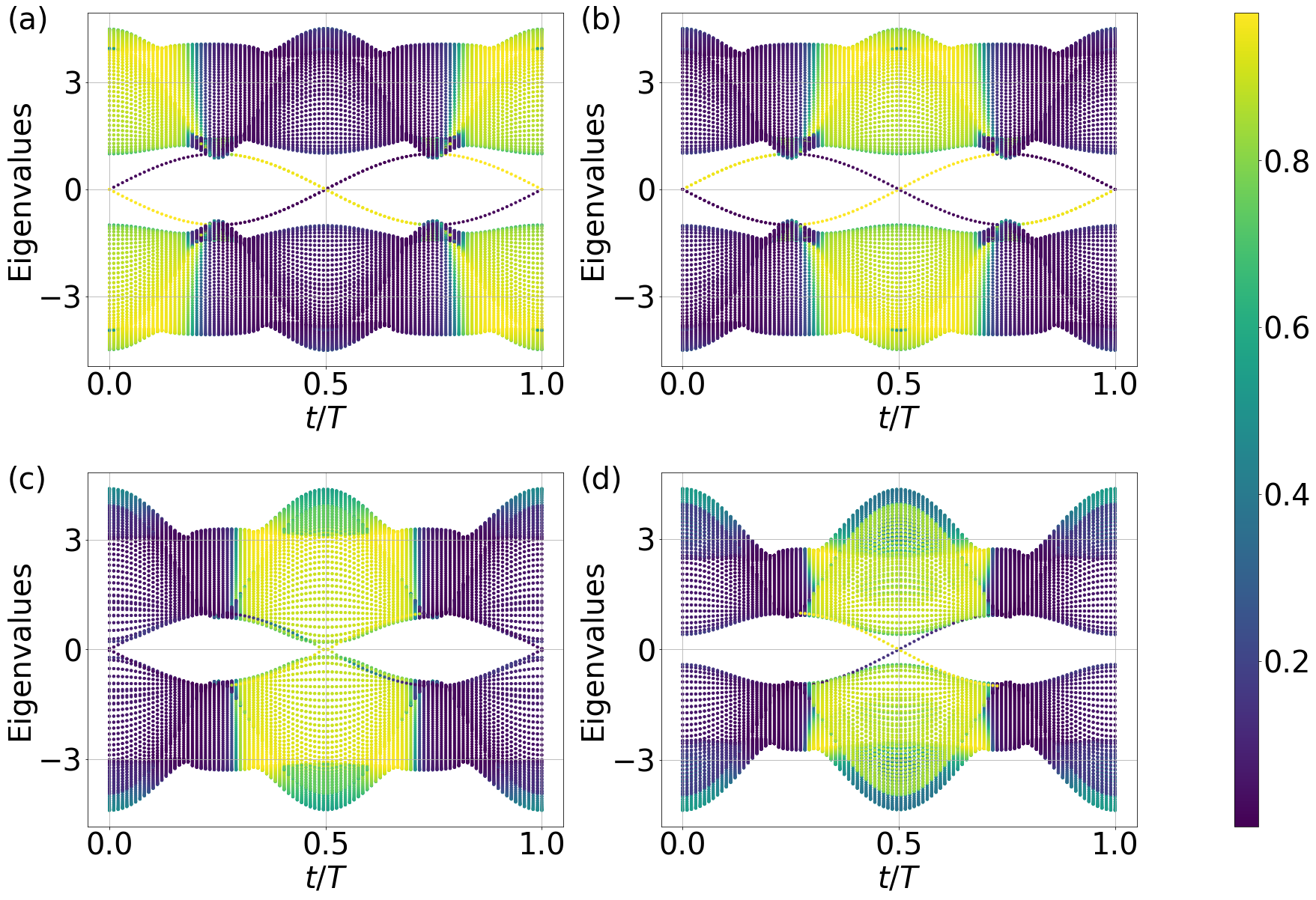}
    \caption{The variation of energy levels, computed from real space time-dependent NH SSHLR model Eq. (1) of the main text under OBC, are shown as a function of time for $(t_0,d_0)=(1,2),(1,-2),(0.6,2)$ and $(0.3,2)$ in (a), (b), (c) and (d) respectively. The color bar represents the average position from which skin effect is clearly visible. The chiral crossings in (a,b,d) denote the existence of the topological phases as a signature of BBC. }
    \label{fig:pump}
\end{figure}

\end{onecolumngrid}


%

\end{document}